\documentclass[preprint2]{aastex63}

\usepackage{graphicx}
\usepackage{rotating}
\usepackage{hyphenat}
\usepackage{footnote}
\usepackage{url}
\usepackage{color,soul}
\usepackage{lineno}
\usepackage{natbib}
\usepackage{multirow}
\usepackage{amsmath}
\usepackage{multirow}
\usepackage{wrapfig}
\usepackage{booktabs}
\usepackage{aas_macros} 
\usepackage{xcolor}

\usepackage{float}
\usepackage{url}
\usepackage{CJK}

\newcounter{supp}[section]

\shorttitle{ExoPop}

\shortauthors{Wang et al.}

\graphicspath{./}

\begin{document}
\begin{CJK*}{UTF8}{gbsn}

\title{A Stellar-Type Dependence in the Rocky and Volatile Composition of Small Exoplanets}

\correspondingauthor{Ji Wang}
\email{wang.12220@osu.edu}

\author[0000-0002-4361-8885]{Ji Wang (王吉)}
\affiliation{Department of Astronomy, The Ohio State University, 140 West 18th Avenue, Columbus Ohio, 43210, USA}

\author[0000-0001-6110-4610]{Caroline Dorn}
\affiliation{Institute for Particle Physics and Astrophysics, ETH Zurich, CH-8093 Zurich, Switzerland}
% \email[show]{dornc@ethz.ch}

\author[0000-0003-4829-7141]{Komal Bali}
\affiliation{Institute for Particle Physics and Astrophysics, ETH Zurich, CH-8093 Zurich, Switzerland}
% \email[show]{kobali@ethz.ch}

% \keywords{Extrasolar rocky planets, Ocean planets, Open source software}

\begin{abstract}
We investigate the rocky and volatile composition of small exoplanets by modeling the population-level distribution of densities using a mixture framework that links interior structure models to observable quantities. We analyze three complementary samples spanning different stellar environments: the Luque \& Pall\'e M-dwarf sample, the DACE M-dwarf sample, and the DACE FGK sample. We consider a log-normal parameterization, which captures a characteristic core mass fraction (CMF) and the intrinsic dispersion to describe a single rocky population. The single rocky population inference suggests a higher CMF for small planets around FGK stars than those around M stars by $\sim$16\% (7-11 $\sigma$ depending on sample selection). We also consider a power-law parameterization, which probes clustering near compositional boundaries at CMF=0.32. The power-law parameterization provides an alternative interpretation: 89.9\% to 97.0\% of the planets around FGK stars are rocky whereas up to 61.6\% (ranging from 4.1\% to 61.6\%) of planets around M stars are rocky. In addition, we find that volatile mass fractions are highly concentrated. For example, to describe the DACE M-dwarf sample using a mixture of rocky, water-rich, and gas-rich planets, we find that more than 99.5\% gaseous planets have an atmospheric mass fraction (AMF) $\lesssim 0.01\%$, and more than 55.4\% (82.7\%) gaseous planets have a water mass fraction (WMF) $\lesssim 0.1\%$ ($\lesssim 1\%$). These results suggest that while volatile-bearing planets are common, their composition and prevalence depend strongly on stellar environment, and their volatile inventories are tightly constrained by formation and evolutionary processes.
% For M-dwarf and solar-type planetary systems, multiple volatile scenarios can reproduce the observed distributions, indicating current data cannot uniquely distinguish the dominant volatile composition. 
\end{abstract}

\section{Introduction}
\label{sect:intro}

Small planets with radii between that of Earth and Neptune dominate the known exoplanet population~\citep{Fressin2013}. Understanding their structural composition is therefore a central problem in exoplanet science. The volatile inventory of a planet, including both its total volatile mass fraction and the chemical nature of those volatiles, strongly influences its bulk density, radius, and long-term evolutionary history~\citep{Owen2013,Madhusudhan2021}. Even modest amounts of volatiles can dramatically alter a planet's structure relative to a purely rocky body~\citep{Lopez2014}. In particular, the presence of hydrogen-helium envelopes or large quantities of water can inflate planetary radii and modify thermal evolution and atmospheric escape processes. Determining the volatile content of small planets is therefore essential for understanding their formation pathways and long-term fates.

Despite extensive observational and theoretical efforts, the nature of the volatile reservoirs in small planets remains debated. One interpretation attributes the low densities observed in many sub-Neptune-size planets to the presence of hydrogen–helium envelopes accreted from the protoplanetary disk~\citep{Rogers_james_2023}. Alternatively, some planets may instead be water-rich bodies~\citep{Burn2024}. These two scenarios imply fundamentally different formation and evolutionary histories. Recent studies have highlighted this debate and proposed new observational diagnostics for distinguishing between water-rich and gas-rich interpretations of the small-planet population \citep[e.g.,][]{Rogers2025, steinmeyer_coupled_2026}. However, current observational constraints have not yet conclusively resolved this question.

Population-level studies have revealed additional structure within the small-planet population. Analyses of mass-radius measurements suggest the existence of distinct classes of planets with different bulk densities, often interpreted as rocky super-Earths and lower-density planets \citep[e.g.,][]{Schulze2024,Dainese2025}. While the existence of multiple populations is now established for small planets around M stars, the physical nature of the low-density population remains uncertain \citep{Dainese2025}. Progress in this area has been aided by improvements in the interior structure models. In particular, the coupling of volatiles between the interior and the atmosphere significantly changes bulk composition interpretations ~\citep[e.g.,][]{dorn_hidden_2021, luo_interior_2024} with clear implications on the planet population levels \citep{rogers2025most}. These models provide a more physically consistent description of volatile-rich planets and allow more reliable inference of their possible compositions. 

{{Another source of uncertainty arises from the choice of planetary samples used for statistical analysis. Different observational datasets have different observational and/or selection biases. Focusing on planets around M stars, a key difference between the planet sample from the~\citet{Luque_2022} and that in \citet{Rogers2024} is the treatment of planets near the radius valley (more details are provided in \S \ref{sec:samples}). For example, seven planets around M stars are excluded by the \citet{Rogers2024} sample because they are classified as sub-Neptunes based on their position relative to the radius valley. In contrast, these planets are retained in the \citet{Luque_2022} sample, leading to a noticeable difference between the two samples in the transition to volatile-rich planets. As new measurements continue to improve the available mass–radius datasets, these complementary samples offer opportunities to cross check results and probe the volatile content of small planets across different environments.}}

{{Most previous studies have focused on matching observed  radii, masses, or densities for individual systems~\citep[e.g., ][]{Santerne18} and often consider only one form of volatile, e.g., gas~\citep{Rogers2023}, or water~\citep{Schulze2024}.}} At population level, it remains uncertain which form of volatile dominates and what the global volatile distribution should be to reproduce the observed density measurements based on high-fidelity structural models.

% In this work we address this question by constructing a generative population model that links physically motivated composition classes to the observable density ratio. Rather than modeling individual planets in isolation, we compare the full empirical distribution of densities to synthetic prior-predictive realizations generated from planet structure models. We consider two parameterizations for the volatile distribution. The first adopts a power-law distribution in composition space, providing a minimal description that tests whether volatile fractions cluster near a physically meaningful boundary. The second uses a log-normal distribution, allowing both a characteristic volatile fraction and an intrinsic spread within each compositional population.

In this work we address this question by constructing planet populations with different volatile types and distributions. The volatile properties are linked to the measurable planet bulk density through a planet interior model. Rather than modeling individual planets in isolation, we compare the full empirical distribution of densities to synthetic predictive realizations based on different assumptions of volatile distributions. We consider two parameterizations. The first adopts a power-law distribution in composition space, providing a minimal description that tests whether volatile fractions cluster near a physically meaningful boundary. The second uses a log-normal distribution, allowing both a characteristic mass fraction and an intrinsic spread.

By fitting these models to observed samples of small planets, we aim to quantify the population-level distribution of volatile content required to reproduce the observed density distribution. We apply the framework to both the \citet{Luque_2022} sample and the more recent compilation from the Data Analysis Center for Exoplanets~\citep[DACE, ][]{otegi2020revisited,parc2024super}, enabling a comparison of volatile distributions across different observational datasets. The inferred distributions can then be interpreted in the context of interior structure and compositions, and planet formation models, allowing us to assess whether particular volatile compositions—such as water worlds or gaseous worlds—are favored or disfavored by the data. 

The paper is organized as follows. Planet Samples are presented in \S \ref{sec:samples}. Planet modeling framework is described in \S \ref{sec:modeling} and the interpolation of the model grid in \S \ref{sec:interpolation}. The mixture framework is described in \S \ref{sec:alpha_inference}. Results and discussions are given in \S \ref{sec:results} and \S \ref{sec:discussion}. Major conclusions can be found in \S \ref{sec:conclusions}. 

\section{Sample Selection}
\label{sec:samples}

We consider three planet samples in this work: the ~\citet{Luque_2022} sample, the DACE-M sample, and the DACE-FGK sample. The ~\citet{Luque_2022} sample follows the same selection procedure as \citet{Schulze2024}, and is therefore identical to the sample presented in that study.

The DACE sample is assembled following \citet{Rogers2024}, based on the curated planet samples from \citet{otegi2020revisited} and \citet{parc2024super}. From this parent sample, we construct two subsamples based on stellar effective temperature. {{The DACE-M sample includes planets orbiting cooler stars with $T_{\rm eff} < 3800~\mathrm{K}$, while the DACE-FGK sample includes planets orbiting hotter stars with $T_{\rm eff} > 3800~\mathrm{K}$ and with equilibrium temperature (T$_{\rm{eq}}$) lower than 1200 K. The latter constraint is due to the upper boundary of T$_{\rm{eq}}$ for our model grid (\S \ref{sec:modeling}). In total, there are 27 planets in the Luque \& Palle M-star sample, 18 in the DACE-M sample, and 12 in the the DACE-M sample. Tables of these samples are provided in \S \ref{app:samples}. }}

{{It is worth discussing the difference between the Luque \& Palle M-star planet sample and the DACE-M sample because of the apparent over-abundance of low-density planets in the Luque \& Palle M-star planet sample. It was interpreted as evidence for water worlds by~\citet{Luque_2022} but this was later challenged by~\citet{Rogers2024}. The sample difference is due to the exclusion of a few low-density planets originally in the Luque \& Palle M-star planet sample from the DACE-M sample. In particular, low-density planets such as K2-146 b \& c, TOI-270 d, TOI-776 b, and L 98-59 d are excluded from the DACE-M sample because their radii are too large to be considered super-Earths. In addition, the DACE-M sample also excludes GJ 3053 b and TOI-1235 b because their radii are above the radius valley. Comparing the two samples, the Luque \& Palle M-star planet sample includes CD-60 8051 b (due to its $T_{\rm eff}=3840$ K higher than the DACE-M cutoff $T_{\rm eff}$), TRAPPIST-1 d \& h, GJ 3053 b \& c, and L 98-59 c, but these planets are not present in the DACE-M sample. The planets that are in the DACE-M sample but not in the Luque \& Palle M-star planet sample are GJ 806 b, LTT 1445 A c, TOI-1468 b, and Wolf 327 b. 
}}

\section{Planet Structural Modeling}
\label{sec:modeling}

The interior structure model employed is based on \citet{Dorn_2015,dorn_generalized_2017} with updates described in \citet{dorn_hidden_2021, luo_interior_2024}.
The underlying forward model consists of three layers: an iron core, a silicate mantle, and a H$_2$-He-H$_2$O or a pure H$_2$O steam atmosphere

% maybe most of the text below should go the the Appendix?
For the solid phase of the iron core, we used the equation of state (EOS) of hexagonal close packed iron \citep{hakim_new_2018,miozzi_new_2020}. 
For the liquid iron phase, we used the EOS from \citep{luo_interior_2024}. 
The silicate mantle is composed of three major species MgO, SiO$_2$, and FeO. 
We modeled the solid phase of the mantle using the thermodynamical model \textsc{Perple\_X} \citep{connolly_geodynamic_2009} for pressures below $\approx 125\,$GPa, while for higher pressures we defined the stable minerals a priori and used their respective EOS from various sources \citep{hemley_constraints_1992,fischer_equation_2011,faik_equation_2018,musella_physical_2019}.
The liquid mantle was modeled as a mixture of Mg$_2$SiO$_4$, SiO$_2$ and FeO, as there is no data for the density of liquid MgO in the required pressure-temperature regime \citep{melosh_hydrocode_2007,faik_equation_2018,ichikawa_ab_2020,stewart_shock_2020}. 
In all cases, the EOS of the different components were mixed using the additive volume law. 
Both the iron core and the silicate mantle were assumed to be adiabatic. 

For gas-rich scenarios, we consider the H$_2$-He-H$_2$O atmosphere layer for which we use the analytic description by \citep{guillot_radiative_2010} and \citep{2014_Jin_planetarypopulation}. It consists of an irradiated layer on top of a non-irradiated layer in radiative-convective equilibrium. We evaluate the Schwarzschild criteria to determine the transition to the convective region.
The water mass fraction (WMF) is given by $Z_\mathrm{env}$, and the hydrogen-helium ratio is set to solar. 
The two components of the atmosphere, H$_2$/He and H$_2$O, were again mixed following the additive volume law. 
We used the EOS by \citep{1995_Saumon_EOS} for H$_2$/He and the ANEOS EOS \citep{1990_thompson_aneos} for H$_2$O. The transit radius of the planet is defined at the radius where the chord optical depth is $\tau_\text{ch}=0.56$. {{We modeled for planets with atmosphere mass fraction (AMF) ranging from 1e-6 to 1e-1 and with temperatures as high as 1200K for this work.}}

The transit radius of a planet with a steam envelope is assumed to be at a pressure of $P_{\rm Transit}=1$ mbar. The thermal profile is assumed to be fully adiabatic, except for pressures less than the pressure at the tropopause (here fixed at 0.1 bar) where we keep an isothermal profile that equals the equilibrium temperature.

Water can be added to the mantle and core melts, depending on its solubility and partitioning behavior, for which we follow \citep{dorn_hidden_2021, luo_interior_2024}. The addition of water reduces the density of the mantle and core melts, for which we follow \citep{bajgain_structure_2015} and decrease the melt density per wt\% water by $0.036$ g cm$^{-3}$. For small WMFs, this reduction is nearly independent of pressure and temperature. The addition of water in core melts lowers the density as described in \citep{luo_interior_2024}. The effect of dissolved water on melting temperature is accounted for. Beyond dissolved water in the deep interior, water can be in solid, supercritical and steam phase, for which we employ the EOS compilation in \citep{haldemann_aqua_2020}. 

\section{Interpolation Function Construction}
\label{sec:interpolation}

To connect the planet structure calculations to the statistical framework described in Section~\ref{sec:alpha_inference}, we construct interpolation functions that map planetary properties to bulk density. These interpolations allow the forward model to evaluate planet bulk density for arbitrary combinations of planetary mass, composition, and thermal state. More specifically, the procedure transforms the raw grids into regular lookup tables in logarithmic coordinates and evaluates these tables using nearest-neighbor interpolation.

\subsection{Structural Modeling Grids}
\label{subsec:grid_sources}

The interpolation functions are constructed from planet structure calculations that tabulate planetary bulk density as a function of planetary mass and thermodynamic state for a set of representative compositions. These grids provide a discrete mapping from planetary mass $m$, equilibrium temperature $T$, and composition to density. The compositions considered in this work correspond to rocky planets dominated by silicate and iron interiors, water-rich planets with substantial volatile layers, and gas-rich planets possessing hydrogen--helium envelopes.

% Because the planet models are computed on discrete grids in $m$ and $T$, direct use of the raw data would require repeated searches through the original structure grids during each likelihood evaluation. To make the statistical inference efficient, the raw model grids are converted into regular interpolation tables before they are used in the forward model.

\subsection{General Structure of the Interpolation Procedure}
\label{subsec:interp_overview}

For each composition, the structure grids are processed through a common sequence of steps. The density of each model is converted into a relative density,
\begin{equation}
\rho_{\rm ratio} =
\frac{\rho}{\rho_0(m)},
\end{equation}
where $\rho_0(m)$ is the density of a reference Earth-like rocky planet evaluated at the same mass. This normalization removes the dominant mass dependence of the bulk density and reduces the dynamic range of the interpolated quantity.

The interpolation coordinates are then transformed to logarithmic variables. For all composition classes, the mass coordinate is represented as $\log m$. For volatile-rich planets, the composition coordinate is also represented logarithmically because the WMF or AMF spans several orders of magnitude. For models with an explicit temperature dependence, the equilibrium temperatures are assigned to discrete temperature bins before the interpolation table is constructed.

A regular grid is then constructed in the relevant interpolation coordinates. Each point on this regular grid is assigned the value of the nearest valid model from the original structure grid. The assigned quantity is the logarithm of the relative density. The final interpolation function is constructed from this populated grid and evaluated using nearest-neighbor interpolation. 
% This procedure avoids fitting a smooth surface through sparsely sampled or irregular regions of the structure grids, while preserving the local values of the original model calculations.

\subsection{Rocky Planet Models}
\label{subsec:rocky_interp}

For rocky planets the density depends primarily on planetary mass and core mass fraction (CMF) and only weakly on temperature over the range considered in this work. Consequently, the rocky composition models are represented using a two-dimensional interpolation function,
\begin{equation}
f_{\mathrm{rocky}}(m, \gamma) = \rho_{\rm ratio},
\end{equation}
where $m$ denotes planetary mass, $\gamma$ is CMF, and $\rho_{\rm ratio}$ represents the resulting bulk density relative to the reference Earth-like rocky with $\gamma=0.32$.

% The rocky interpolation table is constructed in the coordinates
% \begin{equation}
% (\log M,\gamma).
% \end{equation}
% The mass axis is sampled uniformly in $\log M$ over the region covered by the rocky structure grid and the reference baseline model. The composition axis spans the core mass fractions present in the rocky grid. Each point on the regular interpolation table is assigned the logarithmic relative density of the nearest valid rocky model in this coordinate space.

% This construction preserves the local structure of the original rocky grid while allowing the forward model to evaluate rocky-planet densities efficiently. Because the rocky density surface is smooth and depends mainly on mass and core mass fraction, the nearest-neighbor construction reproduces the original grid with very small numerical error.

\subsection{Water-Rich Interior Models}
\label{subsec:water_interp}

Water-rich planets exhibit a stronger dependence on thermal state than rocky planets. This is because the equation of state of water varies significantly with temperature and phase. The corresponding interpolation function for water-rich worlds therefore depends on planetary mass, WMF, and temperature,
\begin{equation}
f_{\mathrm{water}}(m,\gamma,T) = \rho_{\rm ratio}.
\end{equation}
% where $m$ denotes planetary mass, $\gamma$ is water mass fraction, and $T$ represents the characteristic equilibrium temperature.

The water-rich interpolation table is constructed in the coordinates
\begin{equation}
(\log M,\log \gamma,T_{\rm bin}).
\end{equation}
The WMF is represented logarithmically because the grid spans a broad range of volatile abundances. The equilibrium temperature is assigned to a discrete temperature array with a bin size of 50 K. A regular grid is then generated in $\log M$, $\log \gamma$, and $T_{\rm bin}$, and each point is assigned the logarithmic relative density of the nearest valid water-rich model.

% This nearest-neighbor construction avoids imposing an artificial smoothness assumption across phase transitions or sparsely sampled regions of the water-rich grid. The resulting interpolation table therefore preserves the discrete structure of the underlying interior calculations while remaining efficient to evaluate during the statistical inference.

\subsection{Gas-Rich Interior Models}
\label{subsec:gas_interp}

Gas-rich planets also show strong dependence on both planetary mass and thermal state due to the presence of extended hydrogen--helium envelopes. These envelopes introduce large variations in radius and density even for small changes in AMF or temperature. 

The gas-rich interpolation function is defined as
\begin{equation}
f_{\mathrm{gas}}(m,\gamma,T_{\rm bin}) = \rho_{\rm ratio},
\end{equation}
$\gamma$ here is the AMF.

Because the gas-rich grid spans a much larger effective parameter space than the rocky and water-rich grids, the nearest-neighbor approach avoids introducing artificial oscillations in regions where the original grid is sparse, while still capturing the main variation of density with mass, AMF, and temperature.

\subsection{Accuracy of the Interpolator}
\label{subsec:constraints}

We assess the numerical accuracy of the interpolator using pointwise leave-one-out cross-validation by comparing interpolated densities to the original interior-structure grid values. For each composition class, one grid point is removed at a time, the interpolator is reconstructed using the remaining grid points, and the density at the omitted grid point is then predicted. The relative interpolation error is computed as
\begin{equation}
\Delta_{\rho} =
\frac{\rho_{\mathrm{interp}} - \rho_{\mathrm{grid}}}{\rho_{\mathrm{grid}}},
\end{equation}
{{where $\rho_{\mathrm{grid}}$ denotes the density predicted by the original interior structure calculation and $\rho_{\mathrm{interp}}$ is the density returned by the reconstructed interpolation function evaluated at the input parameters of the omitted grid point.}}

\begin{figure*}[t]
\centering
\includegraphics[width=\linewidth]{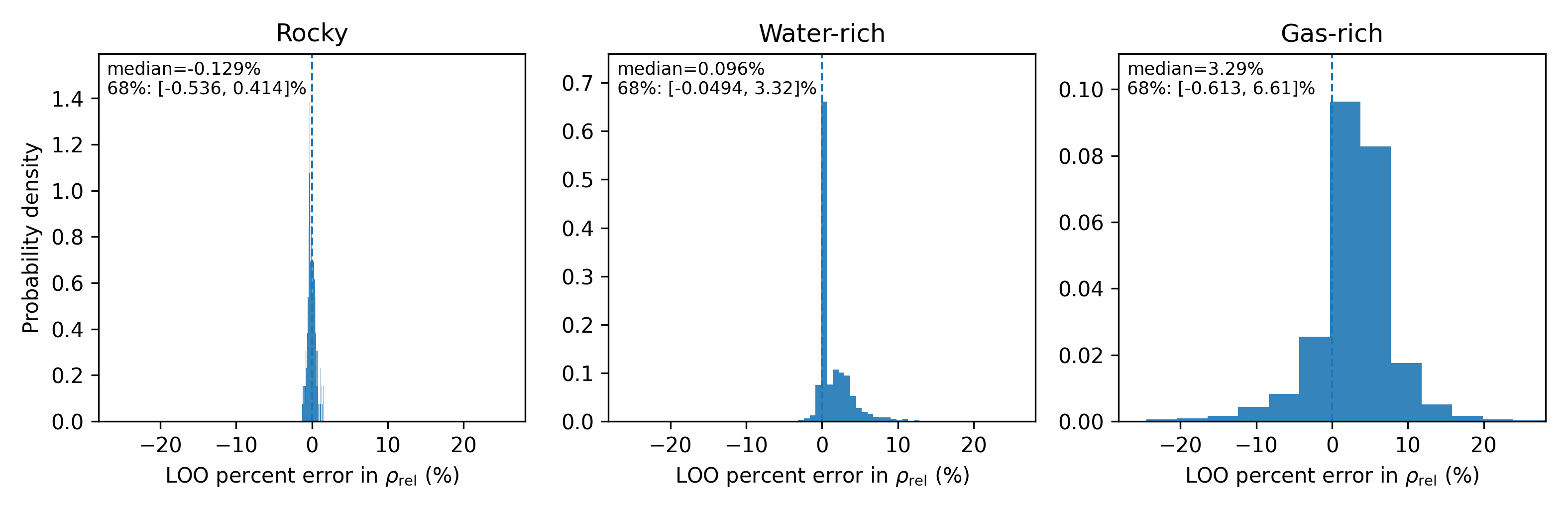}
\caption{Accuracy of the interpolation functions as inferred from the leave-one-out (LOO) test. Each panel shows the probability density distribution of the percent interpolation error,
$\Delta_{\rho} = (\rho_{\mathrm{interp}}-\rho_{\mathrm{grid}})/\rho_{\mathrm{grid}}$, evaluated at the original interior-model grid points. The annotated text indicates the bias and the central $68\%$ range of interpolation errors for each compositional class. In comparison, the average density uncertainty is 13\% for the planet samples around M stars (Table \ref{tab:lp_sample_full} and \ref{tab:dace_m_sample_full}) and 19\% for the DACE FGK sample (Table \ref{tab:dace_fgk_sample_full}). }
\label{fig:interp_accuracy}
\end{figure*}

Figure~\ref{fig:interp_accuracy} shows the distribution of leave-one-out interpolation errors for the three composition classes. Each panel presents the probability density of the percent error in relative density for the rocky, water-rich, and gas-rich interpolator. The annotated intervals indicate the bias and the central $68\%$ range of interpolation errors. {{Overall, the interpolation framework provides a numerically stable mapping between planetary parameters and relative density for all three composition classes. The median signed errors are $-0.13\%$, $0.08\%$, and $3.37\%$ for the rocky, water-rich, and gas-rich interpolator, respectively, while $68\%$ of the absolute interpolation errors are smaller than $0.54\%$, $3.32\%$, and $6.61\%$. These interpolation errors are smaller than the average density uncertainty of $13\%$ for the planet samples around M stars (Tables~\ref{tab:lp_sample_full} and \ref{tab:dace_m_sample_full}) and $19\%$ for the DACE FGK sample (Table~\ref{tab:dace_fgk_sample_full}).}}

\section{Inference of Composition in a Mixture Model}
\label{sec:alpha_inference}

\subsection{Conceptual Framework}

The distribution of $\rho_{\rm ratio}$ contains information about the internal composition and volatile mass fraction for small planets. Recall that $\rho_{\rm ratio}$ is defined in \S \ref{sec:interpolation} as the density ratio between the measured planet bulk density and a reference density for Earth-like rocky planet with $\gamma$=0.32. We construct a model that links the physically motivated composition classes to the observable density ratio. In this framework, the population is described as one single composition or a mixture of distinct compositions, each corresponding to a different compositional regime (e.g., rocky, water-rich, or gas-rich planets).

Each combination of compositions defines how the composition coordinate $\gamma$ is distributed relative to a physically meaningful boundary $\gamma_0$, as well as how $(m,\gamma,T_{\rm eq})$ map to $\rho_{\rm ratio}$ through a forward interior structure model. The key parameters of interest here are the mixture weights, which describe how common each composition is in the observed sample, and the parameters describing the distribution of $\gamma$ within each population. 

We consider two parameterizations for the composition distribution: a power-law model and a log-normal model. In the power-law formulation, a single parameter $\alpha_j$ controls how strongly the distribution clusters toward or spread away from the boundary in composition space, where $j$ is the subscript for a composition or population. In the log-normal formulation, two parameters $(\mu_j,\sigma_j)$ describe the center and spread of the distribution in the logarithmic composition space.

The goal of the inference is therefore not only to reproduce the observed density distribution, but also to investigate what form of composition distribution is required for each candidate population to explain the data.

\subsection{Mixture Over Compositions}

We denote the set of $P$ compositions by $\{\mathcal{P}_j\}_{j=1}^{P}$. The overall population is modeled as a mixture of these compositions with weights $\mathbf{w}=(w_1,\dots,w_P)$ satisfying

\begin{equation}
\sum_{j=1}^{P} w_j = 1, \qquad w_j \ge 0.
\end{equation}

Here $w_j$ represents the fraction of planets belonging to composition $\mathcal{P}_j$. Because the weights lie on the simplex, only $P-1$ parameters are independent. 
% Operationally, the process begins by drawing a population index $j$ from a categorical distribution defined by $\mathbf{w}$, after which the corresponding composition determines the structure of the planet.

\subsection{Latent Mass Distribution and Forward Modeling}

{{Planetary masses are treated as latent variables and are drawn from a sample-specific empirical prior. The prior is an equally weighted mixture of positive-truncated Gaussian kernels centered on the observed planetary masses, with kernel widths set by the corresponding reported mass uncertainties. Thus, the synthetic mass distribution reproduces the observed sample while explicitly accounting for mass measurement errors.}}

For each synthetic planet in the prior-predictive simulation the following steps are performed. A mass $m$ is first drawn from the latent mass prior. A population index $j$ is then drawn according to the mixture weights $\mathbf{w}$. A composition coordinate $\gamma$ is generated according to the composition-specific distribution described below, and an equilibrium temperature $T_{\rm eq}$ is drawn from the empirical temperature distribution of the observed sample. The forward interior structure model associated with composition $\mathcal{P}_j$ is then evaluated to compute

\begin{equation}
\rho_{\rm ratio}=\rho_{\rm ratio}(m,\gamma,T_{\rm eq}).
\end{equation}

Repeating this procedure produces a synthetic realization of the $\rho_{\rm ratio}$ distribution implied by a given parameter set.

\subsection{Power-Law Structure in Composition Space}

The internal structure of each population is controlled through the distribution of the composition coordinate $\gamma$. Instead of modeling $\gamma$ directly, we introduce the distance from the boundary

\begin{equation}
d = |\gamma-\gamma_0|,
\end{equation}
where $\gamma_0$ is set to be 0.32. 

For each composition $\mathcal{P}_j$, the distance $d$ follows a truncated power-law distribution

\begin{equation}
p(d|\alpha_j)=C_j d^{-\alpha_j}, \qquad d_{\min,j} \le d \le d_{\max,j},
\end{equation}

where $\alpha_j$ is the slope parameter, $d_{\min,j}$ is a composition-dependent lower cutoff, $d_{\max,j}$ is determined by the physically allowed range of $\gamma$, and $C_j$ is the normalization constant.

The parameter $\alpha_j$ controls how strongly the distribution concentrates near the boundary. For $\alpha_j=0$ the distribution is uniform in $d$, while $\alpha_j=1$ corresponds to a log-uniform distribution. Larger values of $\alpha_j$ increasingly favor small $d$, producing populations that cluster close to $\gamma_0$.

The mapping from $d$ back to $\gamma$ depends on the composition. For volatile-poor populations,

\begin{equation}
\gamma=\gamma_0+d,
\end{equation}

while for volatile-rich populations,

\begin{equation}
\gamma=\gamma_0-d.
\end{equation}

\subsection{Log-Normal Structure in Composition Space}

As an alternative parameterization, the composition coordinate $\gamma$ may follow a truncated log-normal distribution,

\begin{equation}
\log_{10}\gamma \sim \mathcal{N}(\log_{10}\mu_j,\sigma_j),
\end{equation}

{{where $\log_{10}\mu_j$ and $\sigma_j$ represent the mean and dispersion (in dex) of the distribution in logarithmic composition space for population $\mathcal{P}_j$. }}

Because the physically allowed composition range is finite, the distribution is truncated to the interval permitted by the structure model. The parameters $(\log_{10}\mu_j,\sigma_j)$ therefore encode the characteristic rocky/volatile fraction and the intrinsic spread of each compositional population. Small $\sigma_j$ values correspond to tightly clustered compositions, whereas larger $\sigma_j$ produce broader distributions spanning a wider range of volatile fractions.

\subsection{Prior Assumptions}

The mixture weights are assigned a uniform Dirichlet prior over the simplex. Priors for the composition parameters are summarized in Table~\ref{tab:priors}. The power-law slopes are sampled uniformly in logarithmic space to allow exploration across several orders of magnitude in clustering strength. For the log-normal formulation the mean and dispersion parameters are restricted to physically allowed ranges of $\log_{10}\gamma$. More specifically, $\log_{10}$($\gamma_{\min}$, $\gamma_{\max}$) are $\log_{10}$(0.32, 0.9), $\log_{10}$(0.03, 0.3199), and $\log_{10}$(0.1, 0.319999) for the rocky, water, and gas compositions.
% {{For log-normal dispersion, $\log_{10}$($\sigma_{\min}$, $\sigma_{\max}$) are (-2, -1), (-3, -1), and (-5, -1) for the rocky, water, and gas compositions.}} 
For the volatile compositions, the choice of upper boundaries correspond to the minimum volatile mass fraction allowed by the structure model. Other priors are set to sufficiently wide and therefore uninformative.

\begin{table}
\centering
\scriptsize
\caption{Prior distributions for model parameters.}
\label{tab:priors}
\begin{tabular}{lll}
\hline
Parameter & Prior & Description \\
\hline
$w_j$ & Dirichlet$(1,\dots,1)$ & mixture weights \\
$\alpha_j$ & $\log_{10}\alpha_j \sim \mathcal{U}(-2,0.5)$ & power-law slope \\
$\log_{10}\mu_j$ & $\mathcal{U}(\log_{10}\gamma_{\min},\log_{10}\gamma_{\max})$ & log-normal mean \\
$\log_{10}\sigma_j$ & $\mathcal{U}(\log_{10}\sigma_{\min},\log_{10}\sigma_{\max})$ & log-normal dispersion \\
\hline
\end{tabular}
\end{table}

\subsection{Likelihood Function}

Rather than constructing a per-object likelihood in measurement space, we compare the full empirical distribution of $\rho_{\rm ratio}$ to synthetic realizations. For each parameter set we generate $N_{\rm draw}$ synthetic density ratios {from the population model and the planetary-structure forward model.}

{{Observational uncertainties in $\rho_{\rm ratio}$ are incorporated by convolving the intrinsic synthetic distribution with an empirical error model. For each synthetic planet, we draw an uncertainty $\sigma_{\rho,i}^{\rm mock}$ from the reported $\rho_{\rm ratio}$ uncertainties of the corresponding observational sample and generate

\begin{equation}
\mathbf{
\rho_{{\rm ratio},i}^{\rm mock}
=
\rho_{{\rm ratio},i}^{\rm int}
+
\sigma_{\rho,i}^{\rm mock}\epsilon_i,
}
\end{equation}

where $\rho_{{\rm ratio},i}^{\rm int}$ is the intrinsic forward-model prediction and $\epsilon_i \sim \mathcal{N}(0,1)$. The synthetic distribution therefore inherits the empirical distribution of the reported density-ratio uncertainties. The cumulative distribution function (CDF) comparison is then performed between the observed density ratios and the error-convolved synthetic density ratios.}}

% \textbf{To reduce Monte Carlo fluctuations in the likelihood, the selected uncertainty values and the corresponding Gaussian deviates are generated once and held fixed throughout each nested-sampling run. 

For each parameter set we compute the Cram\'er--von Mises statistic

\begin{equation}
\omega^2 =
\frac{mn}{(m+n)^2}
\sum_{k=1}^{m+n}
\left[F_{\rm data}(z_k)-F_{\rm model}(z_k)\right]^2 ,
\end{equation}

where $m$ and $n$ are the sizes of the observed and simulated samples and $F_{\rm data}$ and $F_{\rm model}$ are the CDFs evaluated on the pooled order statistics $\{z_k\}$. {Here, $F_{\rm model}$ denotes the CDF of the synthetic density ratios after the observational-error convolution.}

The statistic is converted into a pseudo log-likelihood,

\begin{equation}
\log\mathcal{L}=-\frac{\omega^2}{\tau},
\end{equation}

where $\tau$ controls the effective sharpness of the likelihood surface.

\subsection{Posterior Exploration}

The posterior distribution of the model parameters is explored using nested sampling with the \texttt{dynesty} package \citep{speagle20_dynesty}. The dimensionality of the parameter space is $2P-1$ for the power-law model and $3P-1$ for the log-normal model. For each sampled parameter set the forward model generates a realization of the density distribution, which is then evaluated using the likelihood defined above. Posterior summaries, including medians, credible intervals, and maximum a posteriori estimates, are obtained from the nested sampling output. 

\subsection{Choice of Nested Sampling Parameters}

The nested sampling runs are performed using 2000 live points ($n_{\rm live}=2000$) with a stopping criterion of $\Delta\log Z=0.1$. Each likelihood evaluation generates $N_{\rm draw}=10000$ synthetic planets. 

These choices are motivated by stability tests of the likelihood estimator. Repeated evaluations of the likelihood at fixed parameters yield a typical Cram\'er–von Mises scatter of $\sigma_{\rm CvM}\approx2\times10^{-3}$, corresponding to fluctuations in $\log\mathcal{L}$ of order $\sim0.1$ for $\tau=0.02$. This stochastic variation is small compared with the typical variation of the likelihood across the parameter space, ensuring that Monte Carlo noise does not dominate the inference. The number of live points was chosen to ensure adequate exploration of the multimodal posterior surface expected in mixture models while maintaining computational feasibility. Increasing $n_{\rm live}$ beyond this value produces negligible changes in the inferred posterior distributions.
%Komal's comments: did just small grammatical edits so far. In section 4, I like the split of the different functions.

\begin{figure*}[ht!]
\centering
\includegraphics[width=\textwidth]{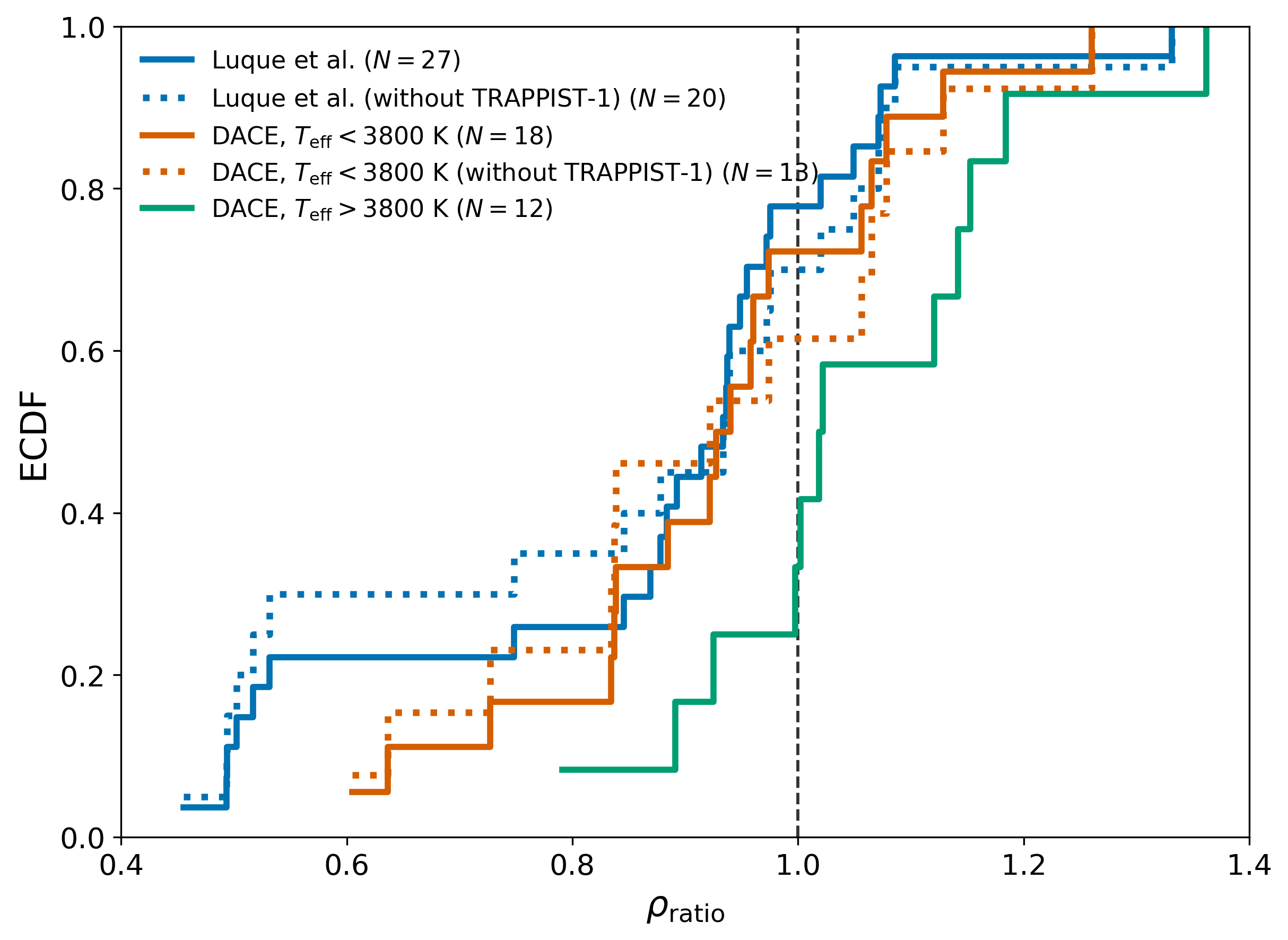}
\caption{
Comparison of the observed density-ratio distributions for the different planet samples considered in this work. The cumulative distribution functions (CDFs) are shown for the Luque \& Palle sample~\citep[][ $N=27$; solid blue]{Luque_2022} , the the Luque \& Palle sample after excluding the TRAPPIST-1 planets ($N=20$; dotted blue), the DACE M-dwarf sample with $T_{\rm eff}<3800$ K ($N=18$; solid orange), the corresponding DACE M-dwarf sample after excluding the TRAPPIST-1 planets ($N=13$; dotted orange), and the DACE FGK sample with $T_{\rm eff}>3800$ K ($N=12$; solid green). Here, $\rho_{\rm ratio}$ is the measured bulk density of each planet normalized by the density expected for a rocky planet of the same mass with a core mass fraction of 0.32. The vertical dashed line indicates the reference value $\rho_{\rm ratio}=1$, corresponding to this nominal rocky composition. The Luque \& Palle sample and DACE-M samples show broadly similar distributions, with both containing a substantial population of planets at $\rho_{\rm ratio}<1$ and extended low-density tails. Removing the TRAPPIST-1 planets does not remove the overall excess of planets below the reference rocky density. In contrast, the DACE FGK sample is systematically shifted toward larger $\rho_{\rm ratio}$, with most planets concentrated near or above $\rho_{\rm ratio}=1$ and comparatively few objects in the low-density tail. This comparison illustrates the differences in the observed $\rho_{\rm ratio}$ distributions among different samples.
}
\label{fig:rho_ratio_data_ecdfs}
\end{figure*}

\section{Results}
\label{sec:results}

The posterior distributions obtained from the aforementioned framework provide constraints on how each candidate population must distribute its rocky and volatile content in order to reproduce the observed density distribution. Below we examine the inferred parameter distributions for the   log-normal and power-law parameterizations and discuss their implications for the composition and formation of small planets. 

\subsection{Describing Planet Populations with a Rocky-Only Composition}
\label{sec:rocky}

\begin{figure}
\centering
\includegraphics[width=\linewidth]{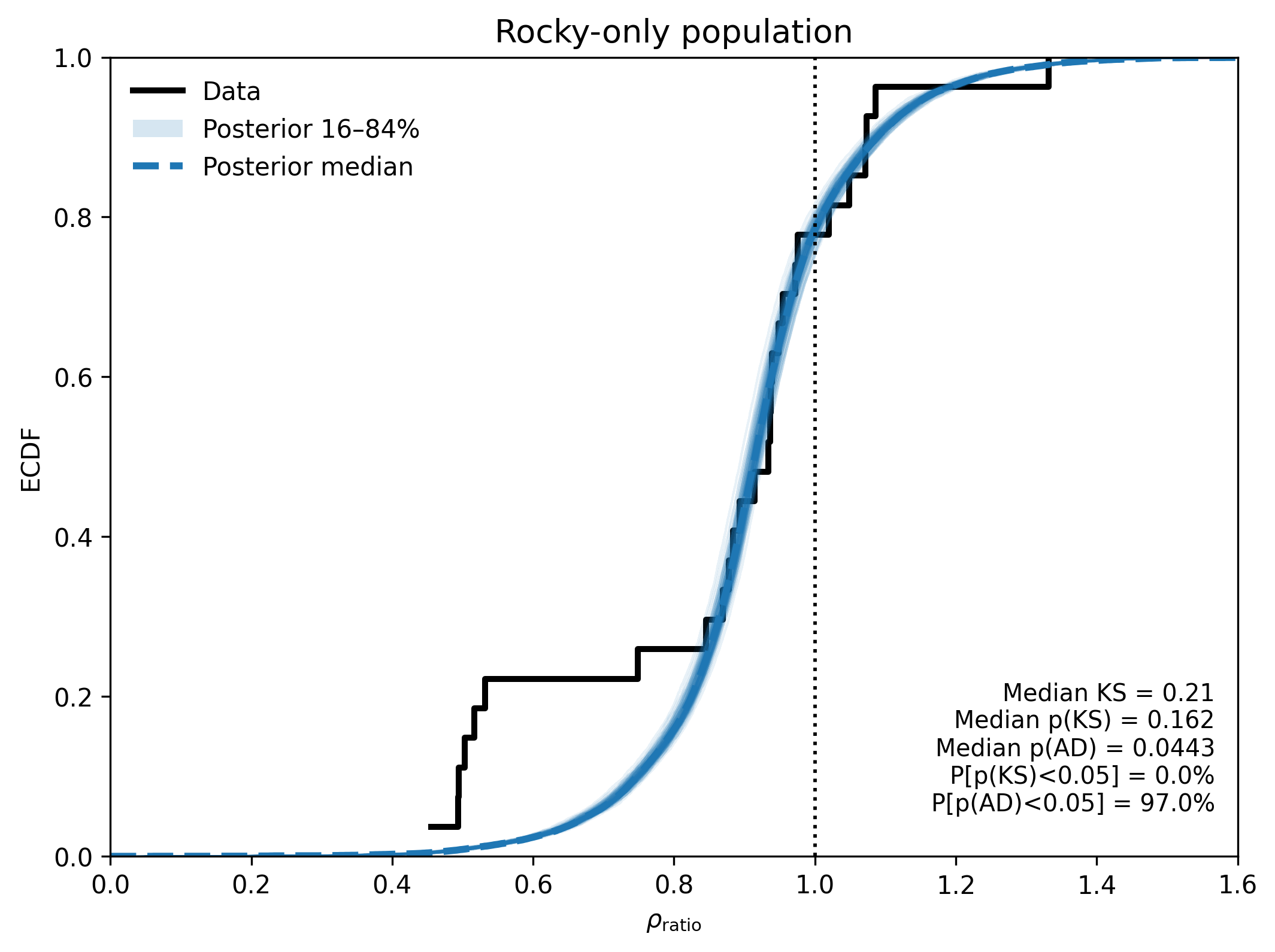}
\caption{Comparison between the observed density-ratio distribution for the {\bf{Luque \& Pall\'e M-dwarf sample}} and posterior realizations using the {\bf{log-normal distribution}} using a {\bf{rocky-only}} composition. The black curve shows the cumulative distribution function (CDF) of the data, while the blue curves represent posterior realizations. The dashed blue line indicates the posterior median CDF. }
\label{fig:luque_cdf_rocky_only}
\end{figure}

{{Figure~\ref{fig:rho_ratio_data_ecdfs} compares the CDFs of $\rho_{\rm ratio}$ for the different planet samples. The Luque \& Palle M-star sample and the DACE M-dwarf sample show broadly similar distributions, including their extended low-density tails. This similarity persists when the TRAPPIST-1 planets are removed from both samples. Because the TRAPPIST-1 system contains seven transiting planets, it can contribute substantial statistical weight to relatively small samples of planets orbiting M stars. The persistence of the same overall CDF structure after removing TRAPPIST-1 therefore suggests that the inferred density distribution is not driven primarily by this single planetary system. In contrast, the DACE FGK sample is clearly shifted toward higher $\rho_{\rm ratio}$ than either M-star sample. This difference motivates a single-rocky-population analysis as a simple means of translating the observed density differences into constraints on the characteristic CMF.

When described as a single rocky population with a log-normal CMF distribution, the DACE FGK sample has a significantly higher characteristic CMF, $\mu_R=36.71^{+1.28}_{-1.27}\%$, than the M-star samples, which have $\mu_R\simeq19$--$22\%$ (see also Table \ref{tab:rocky_only_results}). We estimate the significance of these differences using the difference in $\mu_R$ divided by the quadrature sum of the relevant one-sided uncertainties. The DACE FGK sample differs from the Luque \& Palle M-star sample, the Luque \& Palle sample without TRAPPIST-1, the DACE M-dwarf sample, and the DACE M-dwarf sample without TRAPPIST-1 at approximately $11.3\sigma$, $9.3\sigma$, $9.3\sigma$, and $7.0\sigma$, respectively. The persistence of this difference after removing TRAPPIST-1 highlights a substantial compositional difference between small planets orbiting stars of different spectral types.

The single-rocky-population model nevertheless provides a poorer description of the Luque \& Palle M-star sample (Fig. \ref{fig:luque_cdf_rocky_only}), as indicated by its high rejection rate $f_{\rm AD}$ of $97\%$ ($100\%$ after removing TRAPPIST-1). AD stands for the Anderson--Darling (AD) test. Together with the Kolmogorov--Smirnov (KS) test, we evaluate the consistency between the predictive models and the observed density-ratio distributions. The difference in KS and AD tests highlights the difference in the tail of the density distribution and the importance of using both in robustly testing the null hypothesis that the inferred and the observed distributions are from the same parent distribution. 

This large fraction of AD rejection rate is primarily associated with the low-density concentration near $\rho_{\rm ratio}\sim0.5$, which cannot be readily reproduced by a single narrow rocky population. As discussed in \S~\ref{sec:samples}, this feature likely reflects the inclusion of low-density planets above the radius valley. The discrepancy becomes substantially less important, as discussed in \S \ref{sec:discussion}, once multiple compositions---including rocky, water-rich, and gas-rich populations---are allowed.

Finally, a characteristic CMF of only $\sim20\%$ for small planets around M stars is substantially lower than the CMFs expected from stellar abundance constraints~\citep{Schulze21}. Most of the M-dwarf hosts considered here are in the solar neighborhood and are therefore not expected to exhibit abundance ratios, such as Mg/Fe, sufficiently different from solar values to naturally produce such low rocky CMFs. If these planets instead have intrinsically rocky compositions more similar to those inferred for terrestrial planets around FGK stars~\citep{Dressing2015,Buchhave_2016,Adibekyan_2021,Brinkman2025}, the apparently low CMFs inferred from the M-star samples likely indicate the presence of lower-density volatile material, such as water and/or H/He gas. We therefore consider multi-composition population models in the next section, with the volatile mass fractions described by power-law distributions.

}}

\begin{table*}[t!]
\centering
\caption{The parameter $\delta\mu_R$ denotes the characteristic width of the rocky CMF distribution in linear CMF space and is reported in percentage points. {{Because $\mu=10^{\log_{10}\mu}$, a small dispersion in logarithmic CMF corresponds to a fractional dispersion $\sigma_\mu/\mu \simeq \ln(10)\sigma_{\log_{10}\mu}$. We therefore calculate the characteristic absolute width as $\delta\mu_R \simeq 100\,\mu_R\ln(10)\sigma_{\log_{10}\mu}$, where $\mu_R$ in this expression is expressed as a fraction (e.g., $\mu_R=0.20$ for a CMF of $20\%$).}} The final two columns report the fraction of posterior draws with $p_{\rm KS}<0.05$ and $p_{\rm AD}<0.05$, respectively; smaller fractions indicate a lower rejection rate for the null hypothesis that the model and the data are from the same parent distribution.}
\label{tab:rocky_only_results}
\resizebox{\textwidth}{!}{
\begin{tabular}{lccccc}
\hline\hline
Sample
& $N$
& $\mu_R$ (\%)
& $\delta\mu_R$ (percentage points)
& $f_{\rm KS}$ (\%)
& $f_{\rm AD}$ (\%) \\
\hline

Luque \& Palle M-Star sample
& 27
& $19.84^{+0.79}_{-0.80}$
& $1.50^{+1.78}_{-0.83}$
& 0.00
& 97.00 \\

Luque \& Palle M-Star sample (without TRAPPIST-1)
& 20
& $19.31^{+1.38}_{-1.43}$
& $1.65^{+1.79}_{-0.96}$
& 0.00
& 100.00 \\

DACE M-dwarf sample
& 18
& $22.02^{+0.95}_{-1.00}$
& $1.88^{+2.04}_{-1.09}$
& 0.00
& 0.00 \\

DACE M-dwarf sample (without TRAPPIST-1)
& 13
& $20.86^{+1.88}_{-1.94}$
& $1.83^{+1.93}_{-1.10}$
& 0.00
& 0.00 \\

DACE FGK Sample
& 12
& $36.71^{+1.28}_{-1.27}$
& $2.24^{+2.39}_{-1.10}$
& 0.00
& 0.00 \\

\hline
\end{tabular}
}
\end{table*}

\begin{figure*}[t]
\centering
\includegraphics[width=\textwidth]{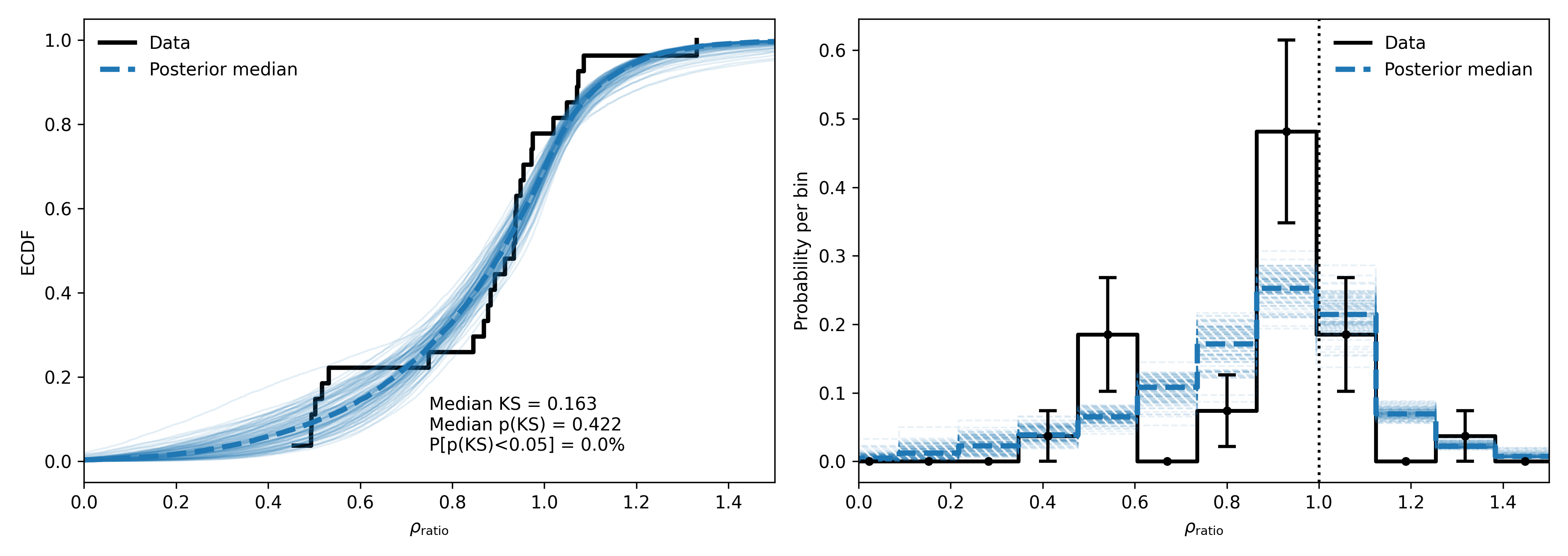}
\caption{
Comparison between observed data~\citep{Luque_2022} and model predictions for the density ratio distribution for the Rocky+Gas population. Results for other cases are in Table \ref{tab:powerlaw_tab} for power-law distributions. Associated figures are in \S \ref{sect:sup_figures}. 
\textit{Left:} Cumulative distribution function (CDF) of the data (black) compared to the posterior distribution (blue), including the median prediction (dashed) and a subset of posterior realizations (light curves). 
\textit{Right:} Histogram representation of the same comparison, showing the binned data with uncertainties and the corresponding posterior distributions. The vertical dashed line indicates the reference density ratio $\rho_{\rm ratio}=1$, corresponding to a rocky planet with CMF=0.32. To assess the quality of the inference, we employ goodness-of-fit tests based on the Kolmogorov--Smirnov (KS) as shown in the figure and Anderson--Darling (AD) statistics which is included in Table \ref{tab:powerlaw_tab}.
}
\label{fig:cdf_hist_comparison}
\end{figure*}

\subsection{Describing Planet Populations with Power-Law Distributions}
\label{sec:powerLaw}

\begin{deluxetable*}{lcccccccc}
\tablecaption{{{Posterior constraints on volatile population mixtures assuming power-law distributions in composition space.}} Results are shown for the Luque \& Palle M-star sample, the DACE M-dwarf sample, and the DACE FGK sample. For the two M-star samples, we additionally show the corresponding constraints after excluding the TRAPPIST-1 planets. The final two columns report the fraction of posterior draws with $p_{\rm KS}<0.05$ and $p_{\rm AD}<0.05$, respectively.\label{tab:powerlaw_tab}}
\tablehead{
\colhead{Sample / Model} &
\multicolumn{2}{c}{Rocky} &
\multicolumn{2}{c}{Water} &
\multicolumn{2}{c}{Gas} &
\colhead{$f_{\rm KS}$} &
\colhead{$f_{\rm AD}$} \\
\colhead{} &
\colhead{$w_R$} &
\colhead{$\alpha_R$} &
\colhead{$w_W$} &
\colhead{$\alpha_W$} &
\colhead{$w_G$} &
\colhead{$\alpha_G$} &
\colhead{(\%)} &
\colhead{(\%)}
}
\startdata
\multicolumn{9}{c}{Luque \& Palle M-star sample} \\
\hline
Rocky+Gas & $0.368^{+0.069}_{-0.067}$ & $2.440^{+0.496}_{-0.649}$ & -- & -- & $0.632^{+0.067}_{-0.069}$ & $2.154^{+0.643}_{-0.587}$ & 0.00 & 0.00 \\
Rocky+Water & $0.041^{+0.087}_{-0.031}$ & $0.665^{+1.715}_{-0.625}$ & $0.959^{+0.031}_{-0.087}$ & $1.067^{+0.062}_{-0.078}$ & -- & -- & 0.00 & 0.00 \\
Rocky+Water+Gas & $0.327^{+0.115}_{-0.228}$ & $2.337^{+0.567}_{-1.159}$ & $0.246^{+0.383}_{-0.167}$ & $0.886^{+0.694}_{-0.820}$ & $0.419^{+0.146}_{-0.255}$ & $1.983^{+0.768}_{-0.702}$ & 0.00 & 0.00 \\
Water+Gas & -- & -- & $0.743^{+0.205}_{-0.207}$ & $1.389^{+0.935}_{-0.230}$ & $0.257^{+0.207}_{-0.205}$ & $1.397^{+0.991}_{-1.275}$ & 0.00 & 0.00 \\
\hline
\multicolumn{9}{c}{Luque \& Palle M-star sample (TRAPPIST-1 removed)} \\
\hline
Rocky+Gas & $0.487^{+0.098}_{-0.129}$ & $2.305^{+0.581}_{-0.760}$ & -- & -- & $0.513^{+0.129}_{-0.098}$ & $1.164^{+0.286}_{-0.163}$ & 0.00 & 0.00 \\
Rocky+Water & $0.402^{+0.186}_{-0.324}$ & $2.158^{+0.695}_{-1.799}$ & $0.598^{+0.324}_{-0.186}$ & $0.685^{+0.306}_{-0.620}$ & -- & -- & 0.00 & 0.00 \\
Rocky+Water+Gas & $0.506^{+0.099}_{-0.174}$ & $2.440^{+0.505}_{-0.735}$ & $0.327^{+0.161}_{-0.151}$ & $0.203^{+0.704}_{-0.175}$ & $0.150^{+0.202}_{-0.111}$ & $1.081^{+1.098}_{-1.007}$ & 0.00 & 0.00 \\
Water+Gas & -- & -- & $0.816^{+0.106}_{-0.105}$ & $1.417^{+0.905}_{-0.274}$ & $0.184^{+0.105}_{-0.106}$ & $0.295^{+0.726}_{-0.263}$ & 0.00 & 0.00 \\
\hline
\multicolumn{9}{c}{DACE M-dwarf sample} \\
\hline
Rocky+Gas & $0.559^{+0.056}_{-0.059}$ & $2.446^{+0.477}_{-0.563}$ & -- & -- & $0.441^{+0.059}_{-0.056}$ & $2.413^{+0.508}_{-0.593}$ & 0.00 & 0.00 \\
Rocky+Water & $0.053^{+0.091}_{-0.039}$ & $0.453^{+1.729}_{-0.419}$ & $0.947^{+0.039}_{-0.091}$ & $1.290^{+0.087}_{-0.091}$ & -- & -- & 0.00 & 0.00 \\
Rocky+Water+Gas & $0.443^{+0.159}_{-0.323}$ & $2.226^{+0.653}_{-1.340}$ & $0.247^{+0.404}_{-0.205}$ & $1.301^{+0.807}_{-1.200}$ & $0.314^{+0.103}_{-0.185}$ & $2.153^{+0.703}_{-0.942}$ & 0.00 & 0.00 \\
Water+Gas & -- & -- & $0.819^{+0.129}_{-0.139}$ & $1.719^{+0.746}_{-0.304}$ & $0.181^{+0.139}_{-0.129}$ & $1.656^{+0.988}_{-1.558}$ & 0.00 & 0.00 \\
\hline
\multicolumn{9}{c}{DACE M-dwarf sample (TRAPPIST-1 removed)} \\
\hline
Rocky+Gas & $0.598^{+0.110}_{-0.100}$ & $2.184^{+0.658}_{-0.678}$ & -- & -- & $0.402^{+0.101}_{-0.110}$ & $2.101^{+0.704}_{-0.727}$ & 0.00 & 0.00 \\
Rocky+Water & $0.179^{+0.511}_{-0.130}$ & $1.386^{+1.249}_{-1.316}$ & $0.821^{+0.130}_{-0.511}$ & $1.207^{+0.144}_{-0.609}$ & -- & -- & 0.00 & 0.00 \\
Rocky+Water+Gas & $0.616^{+0.141}_{-0.288}$ & $2.284^{+0.595}_{-0.879}$ & $0.154^{+0.269}_{-0.110}$ & $0.655^{+1.156}_{-0.614}$ & $0.201^{+0.163}_{-0.136}$ & $1.550^{+1.019}_{-1.460}$ & 0.00 & 0.00 \\
Water+Gas & -- & -- & $0.847^{+0.089}_{-0.137}$ & $1.899^{+0.772}_{-0.429}$ & $0.153^{+0.137}_{-0.089}$ & $0.856^{+1.409}_{-0.812}$ & 0.00 & 0.00 \\
\hline
\multicolumn{9}{c}{DACE FGK sample} \\
\hline
Rocky+Gas & $0.970^{+0.022}_{-0.044}$ & $1.829^{+0.514}_{-0.305}$ & -- & -- & $0.030^{+0.044}_{-0.022}$ & $0.286^{+1.531}_{-0.256}$ & 0.00 & 0.00 \\
Rocky+Water & $0.899^{+0.081}_{-0.355}$ & $1.700^{+0.532}_{-0.488}$ & $0.101^{+0.355}_{-0.081}$ & $1.546^{+0.956}_{-1.455}$ & -- & -- & 0.00 & 0.00 \\
Rocky+Water+Gas & $0.964^{+0.027}_{-0.065}$ & $1.819^{+0.488}_{-0.303}$ & $0.016^{+0.049}_{-0.014}$ & $0.347^{+1.572}_{-0.316}$ & $0.012^{+0.027}_{-0.010}$ & $0.215^{+1.367}_{-0.189}$ & 0.00 & 0.00 \\
Water+Gas & -- & -- & $0.989^{+0.008}_{-0.018}$ & $2.165^{+0.436}_{-0.296}$ & $0.011^{+0.018}_{-0.008}$ & $0.279^{+1.506}_{-0.250}$ & 0.00 & 0.00 \\
\enddata
\end{deluxetable*}

\subsubsection{Rocky Planet Fraction}

The power-law mixture model, despite its simplicity, provides a remarkably good description of the observed planet populations across all samples considered (See Fig. \ref{fig:cdf_hist_comparison} as an example and supplementary figures in \S \ref{sect:sup_figures}). In particular, both the Luque \& Pall\'e M-star sample and the DACE M-dwarf sample are well reproduced by the model, demonstrating that a minimal parameterization is sufficient to capture the key features of the density ratio distribution. For M-dwarf host stars, the inferred population structure is broadly consistent between the two  datasets. Both samples favor a composition in which up to 61.6\% (ranging from 4.1\% to 61.6\%) of planets belong to a rocky population (see the $\omega_R$ rocky weight columns in Table \ref{tab:powerlaw_tab}), while the remaining planets are with inferrable volatile content (either water-rich or gas-dominated). 
% Two exceptions in Table \ref{tab:log_normal_tab} will be discussed in \S \ref{sec:logNormal}.
In comparison, 89.9\% to 97.0\% of the planets in the DACE FGK sample are inferred to be rocky, a $\sim$2$\sigma$ difference from the M star samples. This result, together with the finding in \S \ref{sec:rocky}, suggest a difference between planets orbiting FGK stars and those around M dwarfs, which will be discussed in more details in \S \ref{sec:discussion}.

\begin{figure*}[t]
\centering
\includegraphics[width=0.7\linewidth]{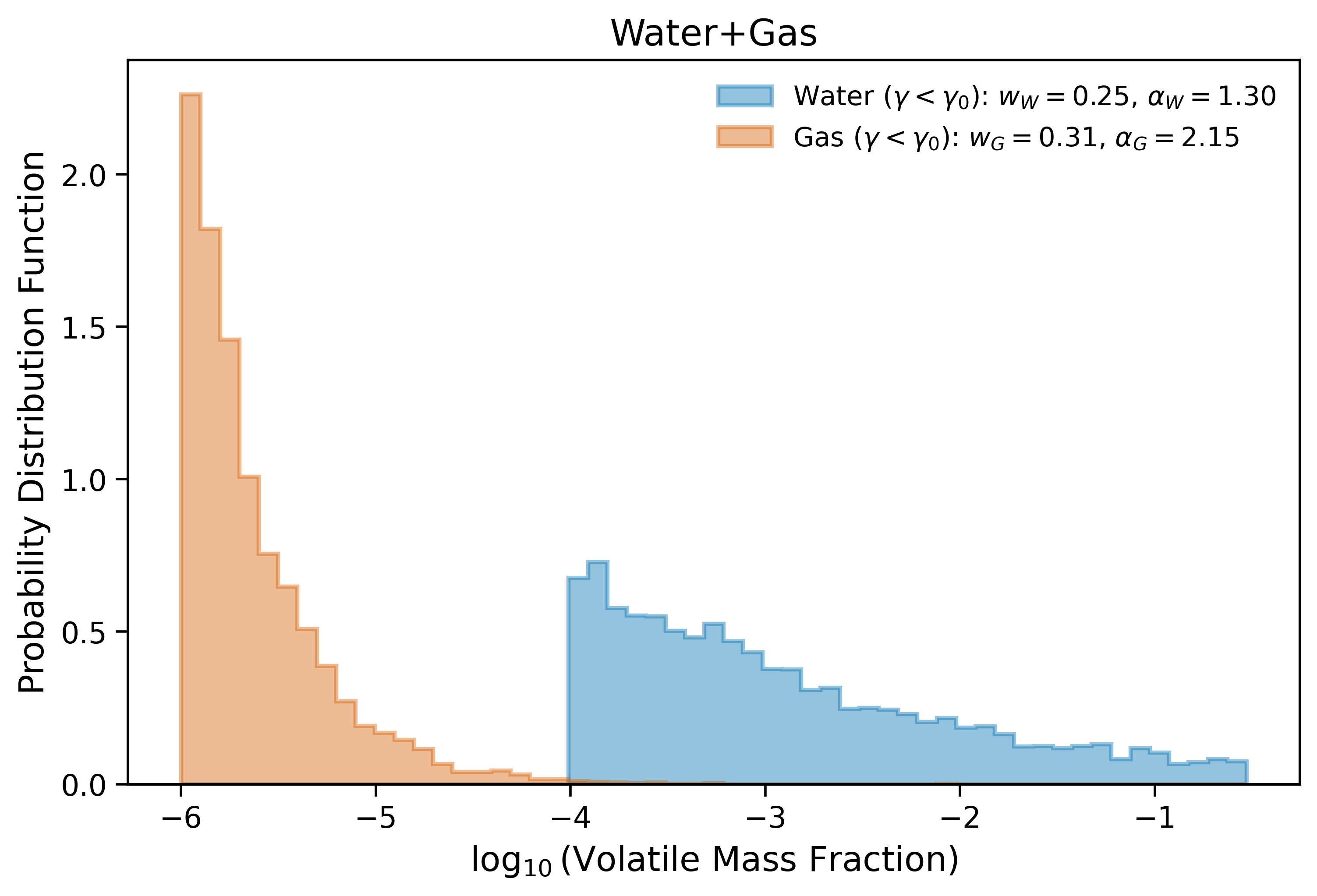}
\caption{
Impact of the power-law index $\alpha$ on the prior distribution of distance from the composition boundary, shown as $\log_{10}|\gamma - \gamma_0|$ or mass fraction for the Rocky+Water+Gas model for the DACE M-dwarf sample. 
The volatile-side distribution ($\gamma < \gamma_0$) with $\alpha > 1$ and favors planets with very small AMFs. In comparison, $\alpha = 1$ is a uniform distribution in $\log_{10}|\gamma - \gamma_0|$ space.
This figure illustrates how the power-law indices control the concentration of planets near the composition boundary. 
Recall that model grid has ($\gamma_{\min}$ and $\gamma_{\max}$) that are (0.32, 0.9), (0.03, 0.3199), and (0.1, 0.319999) for the rocky, water, and gas compositions, and $\gamma_{0}=0.32$. 
% Compositions that are within 0.01, 0.0001, 0.000001 to $\gamma_{0}$ are clipped to $\gamma=0.32$, for rocky, water, and gas compositions. 
}
\label{fig:alpha_impact}
\end{figure*}

\subsubsection{Volatile Distribution}

While a majority of planets around M dwarfs contain volatiles, the amount of these volatiles cannot be arbitrarily large. In particular, the model disfavors scenarios in which substantial gaseous envelopes significantly inflate planetary radii and produce an extended low-density tail. This behavior is most clearly reflected in the inferred power-law index for gas-rich planets. The parameter $\alpha_G$ is greater than unity across all inferences for M-dwarf samples with a sizable gas-rich planet contribution (e.g., $\omega_G$$>$20\%). For inferences with $\omega_G$$>$40\%, $\alpha_G$ ranges from 1.2 to 2.4, indicating a distribution that is strongly weighted toward low AMFs (Fig. \ref{fig:alpha_impact}). In other words, gas-rich planets preferentially cluster at very small envelope fractions: more than 99.5\% gaseous planets have AMFs $\lesssim 0.01\%$ for $\alpha_G$ shown in Fig. \ref{fig:alpha_impact}. This level of AMF is insufficient to substantially increase planetary radii or reduce bulk densities. Similarly, more than 55.4\% (82.7\%) water-rich planets have WMFs $\lesssim 0.1\%$ ($\lesssim 1\%$) for $\alpha_W$ shown in Fig. \ref{fig:alpha_impact}.  

{{Note that the above AMF and WMF constraint comes from one example in Table \ref{tab:powerlaw_tab}. As a more general rule, probability is equal in each dex covered by the model grid for $\alpha$=1, i.e., 20\% per dex and 30\% per dex for the gas and water grid. A higher $\alpha$ value skews the distribution towards smaller values of mass fractions such as the example shown in Fig. \ref{fig:alpha_impact}. To identify the most representative case, we can in principle use model-selection criteria such as the Bayesian Information Criterion (BIC) or the Akaike Information Criterion (AIC). However, both of these metrics require an explicit likelihood function. In our formulation, the likelihood depends on the parameter $\tau$, which describes the intrinsic scatter of the population, so the adopted $\tau$ value inevitably introduces a degree of subjectivity into the likelihood evaluation. We therefore  rely on non-parametric goodness-of-fit statistics, namely the KS and AD tests, to reject unsuccessful cases. }}

\section{Discussion}
\label{sec:discussion}

\subsection{Discussion on Major Findings}
\subsubsection{Larger Fraction (or Higher CMFs) of Rocky Planets Around Solar-Type Stars than M Stars}

{{Our single rocky population inference (\S \ref{sec:rocky}) suggests a CMF difference $\sim$16\% (7-11 $\sigma$) between small planets around M stars and around FGK stars. Alternatively, the CMF difference can translate into fractional difference in rocky planets vs. planets with volatiles such as water and gas. In our mixture model analyses (\S \ref{sec:powerLaw}), 89.9\% to 97.0\% of the planets in the DACE FGK sample are inferred to be rocky whereas M-star samples favor a composition mixture in which up to 61.6\% (ranging from 4.1\% to 61.6\%) of planets belong to a rocky population (Table \ref{tab:powerlaw_tab}).}}

This statistics is not inconsistent with previous works that find that the occurrence rate of small rocky planets are more prevalent around M stars~\citep{Dressing2015,Mignon2025} than around solar-type stars~\citep{Petigura2013PNAS}. Our finding here suggests that among the existing small planet populations, the fraction of rocky planets is higher for solar-type stars than M stars. 

To understand this, we discuss our model assumption and limitation, the difference of the samples, and the underlying formation processes. 

Our mixture model analyses infer a lower rocky planet fraction around M stars because a large fraction of these planets have a CMF or $\gamma<0.32$ and therefore lower-than-unity density ratios by our definition: rocky planets in our mixture model have $\gamma\geq0.32$. Additional volatile content is therefore required to explain the lower-than-unity density ratios. Alternatively, the planets around M stars can have an overall lower CMFs than those around FGK stars as shown in \S \ref{sec:powerLaw}. This would explain the density difference between the M and FGK populations. However, it also suggests that planets around M stars do not reflect the stellar refractory ratios, which is not supported by current observational evidence~\citep{Schulze21,Adibekyan_2021}.  

In terms of sample selection, the solar-type small planet sample have a median of 4.7 M$_{\rm{Earth}}$ whereas the M-star samples have medians of 1.9 and 2.3 M$_{\rm{Earth}}$. For more massive planets around solar-type stars in our sample, some of them have already been shown to have rocky planet composition~\citep{Dressing2015,Buchhave_2016,Adibekyan_2021,Brinkman2025}. On the other hand, less massive planets around M stars, in particular planets around Trappist-1, are shown to have a lower density than a typical rocky planet~\citep{Agol2021}. Therefore, the two samples by construction have a systemic density offset. However, we investigate the sensitivity of our conclusions to the inclusion of Trappist-1 planets and find that our conclusions do not change significantly by the inclusion or exclusion of Trappist-1 planets.    

% The systemic density offset may simply stem from the inclusion of Trappist-1 planets. Excluding the Trappist-1 planets, rocky planets fraction would increases by 23\% and 28\% for the Luque \& Pall\'e sample and the DACE M sample, making it more comparable to the rocky planet fraction for the DACE FGK sample. 

The difference in fraction of rocky planets (or CMFs) can also stem from different formation processes and stellar environments. For example, pebble accretion and migration would play a critical role in determining a planet composition~\citep{Unterborn2018}. If Trappist-1 system is a representative case, then the difference can be attributed to the formation processes. In addition, atmospheric evolution plays an equally important role. The median temperature for the solar planet sample is $\sim$1000 K as compared to the median temperature of 500-600 K for the M star sample. The lower planet equilibrium temperature may facilitate the retention of the planet volatile content, although the prolonged XUV history poses challenges. If the spectral-type dependent rocky planet fraction (or density difference) is confirmed, it would shed light on the timing of rocky planet formation, volatile retention, and volatile transportation in low-mass disks~\citep[e.g., ][]{Xie2023}.    

\subsubsection{Narrow Distribution of Volatile Mass Fractions}

Another major finding is that the volatile fractional mass distribution is highly concentrated. In the power-law case where planets with gaseous envelopes are inferred to dominate the low-density population. The power-law index is consistently above 1.0 for mixture inference with significant fraction of gaseous planets, indicating AMF is highly concentrated toward the rocky-gas edge at CMF=0.32. Indeed, for a power-law $\alpha$ of 2.1 as shown in Fig. \ref{fig:alpha_impact}, only $<$0.5\% of the planets can reach an AMF above 0.01\% to significantly puff up the planet to reduce the density~\citep{Lopez2014}. Similarly, for a power-law $\alpha$ of 1.3 as shown in Fig. \ref{fig:alpha_impact}, only $<$17\% of the planets can reach a WMF above 1\%. 

The highly concentrated volatile distribution shed light on the planet formation and subsequent evolution processes. The low AMF value is likely a result of atmospheric loss~\citep{Owen2024}. Planets that retain a higher AMF are considered sub-Neptunes and therefore not selected into the samples that we consider in this work. The finding that only $<$17\% of the planets can reach a WMF above 1\% (for $\alpha$=1.3) is broadly consistent with theoretical expectations that super-Earth water worlds may contain up to a few percent water by mass, largely independent of their original accretion pathways~\citep{Rogers2024}. The relatively small water inventories suggest that extremely water-rich planets, such as the proposed ``Hycean'' worlds~\citep{Madhusudhan2021}, are unlikely to represent a dominant population within the current small planet sample.

\subsection{Comparison to Previous Findings}

Several recent studies have investigated the composition distribution of small planets using Gaussian mixture models (GMMs) and hierarchical Bayesian Method (HBM). In particular, \citet{Schulze2024} and \citet{Dainese2025} applied GMM-based analyses and reported statistical evidence for multiple compositional populations. These works typically evaluate different population models using Bayesian evidence and conclude that models containing more than one population are preferred over single-population scenarios. {{In particular, \citet{Schulze2024} analyzed the same Luque \& Palle M-dwarf sample as in Table \ref{tab:lp_sample_full}---which applied a cut to exclude some sub-Neptunes---with a mixture model of rocky and water worlds. They found evidence of two distinct populations, which supported the conclusion in~\citet{Luque_2022}.~\citet{Dainese2025} considered a mixture model of rocky, water-rich, and gas-rich planets. They found while their sample (planets around stars cooler than 4000 K and including sub-Neptunes) was best described by two distinct planet populations, there was no clear evidence supporting the existence of water worlds. }}

When our framework is applied in a similar GMM+HBM context, we likewise find that models containing multiple populations are generally favored over a single-population model. However, there are several important caveats that complicate a direct comparison between the results of these studies and those presented here.

First, Bayesian evidence is intrinsically sensitive to the choice of priors. Different model comparisons in the literature involve different prior assumptions, including the allowed parameter ranges and the functional forms of the priors. For example, the comparison between a rocky-only model and a rocky+water model in \citet{Schulze2024}, and the comparison between rocky+gas and rocky+water+gas models in \citet{Dainese2025}, involve distinct prior choices for both mixture weights and distribution parameters. Because the Bayesian evidence integrates the likelihood over the full prior volume, differences in prior ranges or parameterizations can lead to significant changes in the computed evidence. As a result, the relative preference for one model over another may partly reflect differences in prior assumptions rather than purely the constraining power of the data.

Second, the effectiveness of the GMM+HBM framework depends on the dimensionality of the parameter space being modeled. In previous works such as \citet{Schulze2024}, the Gaussian mixture is applied primarily in one dimension, typically the compositional parameter $\gamma$. In this case the data provide sufficient leverage to constrain the mixture components and their relative weights. However, when extending the model to two dimensions—for example by introducing correlations between $\mu$ and planetary mass—the inferred posterior distributions tend to revert toward the prior distributions. This behavior indicates that the current data do not provide strong constraints on the additional parameters introduced in the two-dimensional Gaussian case. Consequently, the statistical preference for specific mixture components becomes much weaker once the model complexity increases beyond the one-dimensional formulation.

% Thirdly, alternative model-selection criteria such as the Bayesian Information Criterion (BIC) or the Akaike Information Criterion (AIC) could in principle be used to compare models with different numbers of populations. However, both of these metrics require an explicit likelihood function. In our formulation, the likelihood depends on the parameter $\tau$, which describes the intrinsic scatter of the population. Although $\tau$ can be motivated physically, its adopted value inevitably introduces a degree of subjectivity into the likelihood evaluation. To avoid this dependence on a particular choice of $\tau$, we instead rely on non-parametric goodness-of-fit statistics, namely the KS and AD tests, which compare the predicted and observed cumulative distributions directly.

{{Finally, the similarity between the results for the Luque \& Pall\'e sample and the DACE M-dwarf sample  suggests that the previously reported density enhancement around $\rho \sim 0.5$ may not be as pronounced as inferred in earlier studies~\citep{Luque_2022,Schulze2024}, or at least cannot be readily reproduced by the power-law parametrization. One might expect that a sharply peaked log-normal distribution of volatile content would naturally reproduce the density enhancement around $\rho \sim 0.5$. However, in practice, the resulting density distribution is significantly smoothed by the intrinsic spread in planetary mass and equilibrium temperature. As a result, even a narrow underlying volatile distribution does not translate into a correspondingly sharp feature in the observed density distribution. This suggests that the narrow concentration at $\rho \sim 0.5$ is not due to a sharp distribution of volatile content. Alternatively, the sharp feature in the density distribution may be a statistical fluke due to small number statistics, or requires a more sophisticated volatile distribution to explain. }}

% Using these statistics, we find that the current data do not provide strong statistical preference for any specific volatile-composition combination. While models with multiple populations are generally consistent with the observations, the KS and AD tests indicate that several different composition scenarios remain viable given the available data.

\section{Conclusions}
\label{sec:conclusions}

In this work, we developed a population model that links interior structure models to the observed density-ratio distribution of small exoplanets through mixture modeling. By applying this approach to three complementary samples (Luque \& Pall\'e, DACE-M, and DACE-FGK) and exploring both power-law and log-normal parameterizations of volatile content, we derive the following conclusions:

\begin{enumerate}

\item \textbf{Stellar environment shapes planet composition.} 
{{We find that planets orbiting solar-type (FGK) stars are dominated by rocky compositions: 89.9\% to 97.0\% of the planets in the DACE FGK sample are inferred to be rocky whereas M-star samples favor a composition mixture in which up to 61.6\% (ranging from 4.1\% to 61.6\%) of planets belong to a rocky population (Table \ref{tab:powerlaw_tab}). Alternatively, the difference can also be explained by a CMF difference $\sim$16\% (7-11 $\sigma$) between small planets around M stars and around FGK stars (\S \ref{sec:rocky}). If confirmed, this may suggest a fundamental difference in the underlying small planet populations around stars of different spectral types. More discussions can be found in \S \ref{sec:discussion}. }}

\item \textbf{Volatile mass fractions are tightly constrained.} 
Volatile inventories are highly concentrated. In the power-law model, gas-rich planets usally exhibit $\alpha_G > 1$, implying strong clustering toward extremely low AMFs. For example, Fig. \ref{fig:alpha_impact} shows that only $<$0.5\% of the planets can reach an AMF above 0.01\%, and only $<$17\% of the planets can reach a WMF above 1\% for a mixture model of rocky, water-rich, and gas-rich planets to describe the DACE M sample. This result puts tight constraints on formation (e.g., location and migration) and evolution (e.g., accretion and atmospheric loss) of super-Earths as discussed in \S \ref{sec:discussion}.

\item \textbf{Degeneracy of volatile composition.} 
For all planet samples, there are multiple volatile scenarios that can reasonably reproduce the observed density distributions and {{are not rejected by}} the KS and AD tests. This indicates that current data do not provide sufficient statistical leverage to distinguish whether the low-density population is dominated by water-rich planets, gas-rich planets, or a mixture of both.

% \item \textbf{Multiple compositional populations are required.} 
% Across all samples, the observed density distributions cannot be explained by a single rocky population unless invoking unrealistically low-density rocky compositions. Instead, the data favor multiple compositional branches, consistent with previous studies, although the exact nature of the volatile-rich branch remains uncertain.

\end{enumerate}

% Overall, our results suggest that while volatile-bearing planets are common, their compositions are tightly regulated and depend strongly on stellar environment. The low inferred volatile mass fractions are consistent with efficient atmospheric loss and/or formation pathways that limit volatile retention. Future observations and better characterized planet masses and radii will be essential for breaking the degeneracy between water-rich and gas-rich interpretations and for refining our understanding of the origin of small exoplanets.

To sum up, bulk density, from which CMF, rocky planet fraction, and volatile mass distribution are inferred, provides complementary population-level tests of planet formation models. Occurrence rates alone, even when resolved by stellar spectral type, offer only a partial view of planet formation~\citep{Mulders2015,Chachan2023,Pan2025}. The results in this work therefore add a new axis for comparison across stellar hosts: not simply how often small planets occur, but how their compositions differ.

The contrast between planets around FGK and M stars is especially revealing. If small planets around M dwarfs are systematically lower density, then existing models and their future expansions~\citep{Schoonenberg2019, Coleman2019,Ogihara2022} must explain whether this reflects a lower fraction of intrinsically rocky planets, enhanced volatile retention, different CMFs, or a combination of these effects. Such compositional differences extend the study of planet formation beyond occurrence demographics alone, and place new constraints on models of accretion, migration, atmospheric loss, and their dependence on stellar spectral type.

More broadly, our inference that volatile-bearing planets are common, yet occupy a limited range of volatile mass fractions, points to a regulating process in the assembly or evolution of small planets. Whatever its origin, this apparent narrow range suggests that volatile enrichment is not arbitrary, but shaped by underlying physical mechanisms that operate across stellar environments. Composition should therefore be as fundamental a demographic observable as occurrence itself in uncovering the origin of small exoplanets.

\section{Acknowledgments}
The authors thank the anonymous referee for insightful comments and suggestions that significantly improved the manuscript. JW acknowledges the support by the National Science Foundation under Grant No. 2143400. JW acknowledges the support through the Humboldt Research Fellowship for Experienced Researchers to carry out the research in Germany. This work is made possible by a generous support by Department of Physics at ETH Zurich through a Visiting Professorship. C.D. acknowledges support from the Swiss National Science Foundation under grant TMSGI2\_211313. This work has been carried out within the framework of the NCCR PlanetS supported by the Swiss National Science Foundation under grant 51NF40\_205606. JW thanks James Rogers for insightful discussions and supplying the DACE sample.

\newpage

\begingroup
\renewcommand{\section}[2]{}%

\bibliography{biblio.bib}
\endgroup

\newpage
\section*{Appendix}

\subsection{Samples}
\label{app:samples}

{{This section provides tables for the three samples used in the analyses. }}

\begin{deluxetable*}{lcccccc}
\tablewidth{0pt}
\tablecaption{\citet{Luque_2022} Planet Sample\label{tab:lp_sample_full}}
\tablehead{
\colhead{Index} & \colhead{Planet} & \colhead{$M~(M_\oplus)$} & \colhead{$R~(R_\oplus)$} & \colhead{$\rho/\rho_{\rm rock}$} & \colhead{$T_{\rm eff}$ (K)} & \colhead{$T_{\rm eq}$ (K)}
}
\startlongtable
\startdata
1 & TRAPPIST-1 g & $1.32 \pm 0.04$ & $1.13 \pm 0.02$ & $0.88 \pm 0.04$ & 2566 & 197 \\
2 & TRAPPIST-1 f & $1.04 \pm 0.03$ & $1.04 \pm 0.01$ & $0.91 \pm 0.04$ & 2566 & 218 \\
3 & TRAPPIST-1 d & $0.39 \pm 0.01$ & $0.79 \pm 0.01$ & $0.89 \pm 0.05$ & 2566 & 300 \\
4 & TRAPPIST-1 e & $0.69 \pm 0.02$ & $0.92 \pm 0.01$ & $0.94 \pm 0.05$ & 2566 & 251 \\
5 & TRAPPIST-1 c & $1.31 \pm 0.06$ & $1.10 \pm 0.01$ & $0.95 \pm 0.05$ & 2566 & 341 \\
6 & TRAPPIST-1 b & $1.37 \pm 0.07$ & $1.12 \pm 0.01$ & $0.95 \pm 0.06$ & 2566 & 399 \\
7 & TRAPPIST-1 h & $0.33 \pm 0.02$ & $0.76 \pm 0.01$ & $0.87 \pm 0.07$ & 2566 & 171 \\
8 & GJ 3053 b & $6.38 \pm 0.45$ & $1.64 \pm 0.05$ & $1.05 \pm 0.12$ & 3096 & 226 \\
9 & GJ 3053 c & $1.76 \pm 0.17$ & $1.17 \pm 0.04$ & $1.02 \pm 0.14$ & 3096 & 422 \\
10 & GJ 1132 b & $1.66 \pm 0.23$ & $1.13 \pm 0.06$ & $1.07 \pm 0.22$ & 3229 & 584 \\
11 & GJ 486 b & $2.82 \pm 0.12$ & $1.30 \pm 0.07$ & $1.09 \pm 0.17$ & 3317 & 702 \\
12 & GJ 3473 b & $1.86 \pm 0.30$ & $1.26 \pm 0.05$ & $0.85 \pm 0.17$ & 3347 & 768 \\
13 & LTT 3780 b & $2.47 \pm 0.24$ & $1.32 \pm 0.06$ & $0.93 \pm 0.16$ & 3358 & 911 \\
14 & LHS 1478 b & $2.33 \pm 0.20$ & $1.24 \pm 0.05$ & $1.07 \pm 0.16$ & 3381 & 600 \\
15 & K2-146 b & $5.77 \pm 0.18$ & $2.05 \pm 0.06$ & $0.49 \pm 0.05$ & 3385 & 534 \\
16 & K2-146 c & $7.49 \pm 0.24$ & $2.19 \pm 0.07$ & $0.49 \pm 0.05$ & 3385 & 391 \\
17 & L 98-59 c & $2.42 \pm 0.34$ & $1.34 \pm 0.07$ & $0.88 \pm 0.19$ & 3412 & 526 \\
18 & L 98-59 d & $2.31 \pm 0.46$ & $1.58 \pm 0.08$ & $0.52 \pm 0.13$ & 3412 & 416 \\
19 & GJ 1252 b & $1.32 \pm 0.28$ & $1.19 \pm 0.07$ & $0.75 \pm 0.21$ & 3458 & 1089 \\
20 & TOI-1634 b & $7.57 \pm 0.71$ & $1.77 \pm 0.08$ & $0.94 \pm 0.15$ & 3472 & 924 \\
21 & GJ 357 b & $1.84 \pm 0.31$ & $1.22 \pm 0.08$ & $0.94 \pm 0.25$ & 3505 & 527 \\
22 & TOI-270 d & $4.78 \pm 0.46$ & $2.00 \pm 0.07$ & $0.46 \pm 0.06$ & 3506 & 383 \\
23 & TOI-270 b & $1.58 \pm 0.26$ & $1.15 \pm 0.05$ & $0.97 \pm 0.20$ & 3506 & 582 \\
24 & TOI-1685 b & $3.09 \pm 0.58$ & $1.70 \pm 0.07$ & $0.53 \pm 0.12$ & 3575 & 1067 \\
25 & TOI-776 b & $4.00 \pm 0.90$ & $1.85 \pm 0.13$ & $0.50 \pm 0.15$ & 3725 & 520 \\
26 & CD-60 8051 b & $4.60 \pm 0.56$ & $1.39 \pm 0.09$ & $1.33 \pm 0.31$ & 3800 & 982 \\
27 & TOI-1235 b & $6.69 \pm 0.68$ & $1.69 \pm 0.08$ & $0.98 \pm 0.17$ & 3997 & 775 \\
\enddata
% \tablecomments{Mass, radius, and relative density are listed with symmetric uncertainties and are therefore reported using $\pm$ notation. Equilibrium temperatures are given as central values because no temperature uncertainties were provided in the input tables.}
\end{deluxetable*}

\begin{deluxetable*}{lcccccc}
\tablewidth{0pt}
\tablecaption{DACE-M Planet Sample ($T_{\rm eff}<3800$ K)\label{tab:dace_m_sample_full}}
\tablehead{
\colhead{Index} & \colhead{Planet} & \colhead{$M~(M_\oplus)$} & \colhead{$R~(R_\oplus)$} & \colhead{$\rho/\rho_{\rm rock}$} & \colhead{$T_{\rm eff}$ (K)} & \colhead{$T_{\rm eq}$ (K)}
}
\startlongtable
\startdata
1 & TRAPPIST-1 g & $1.33 \pm 0.04$ & $1.13 \pm 0.01$ & $0.88 \pm 0.04$ & 2566 & 198 \\
2 & TRAPPIST-1 f & $1.05 \pm 0.03$ & $1.04 \pm 0.01$ & $0.93 \pm 0.04$ & 2566 & 218 \\
3 & TRAPPIST-1 e & $0.70 \pm 0.02$ & $0.92 \pm 0.01$ & $0.96 \pm 0.05$ & 2566 & 251 \\
4 & TRAPPIST-1 c & $1.32 \pm 0.06$ & $1.10 \pm 0.01$ & $0.96 \pm 0.05$ & 2566 & 341 \\
5 & TRAPPIST-1 b & $1.38 \pm 0.07$ & $1.12 \pm 0.01$ & $0.94 \pm 0.06$ & 2566 & 399 \\
6 & GJ 1132 b & $1.67 \pm 0.23$ & $1.13 \pm 0.06$ & $1.07 \pm 0.22$ & 3270 & 584 \\
7 & LTT 1445 A c & $1.55 \pm 0.20$ & $1.14 \pm 0.06$ & $0.97 \pm 0.19$ & 3340 & 513 \\
8 & GJ 486 b & $2.84 \pm 0.12$ & $1.30 \pm 0.07$ & $1.08 \pm 0.17$ & 3340 & 703 \\
9 & GJ 3473 b & $1.87 \pm 0.30$ & $1.27 \pm 0.04$ & $0.83 \pm 0.16$ & 3347 & 768 \\
10 & LTT 3780 b & $2.36 \pm 0.24$ & $1.35 \pm 0.06$ & $0.84 \pm 0.13$ & 3360 & 911 \\
11 & LHS 1478 b & $2.35 \pm 0.20$ & $1.24 \pm 0.05$ & $1.06 \pm 0.16$ & 3381 & 600 \\
12 & TOI-1685 b & $3.81 \pm 0.63$ & $1.70 \pm 0.07$ & $0.61 \pm 0.12$ & 3434 & 1067 \\
13 & TOI-1468 b & $3.24 \pm 0.24$ & $1.28 \pm 0.03$ & $1.26 \pm 0.14$ & 3496 & 681 \\
14 & GJ 357 b & $1.86 \pm 0.31$ & $1.22 \pm 0.08$ & $0.92 \pm 0.24$ & 3505 & 527 \\
15 & TOI-270 b & $1.59 \pm 0.26$ & $1.21 \pm 0.03$ & $0.84 \pm 0.15$ & 3506 & 583 \\
16 & Wolf 327 b & $2.55 \pm 0.46$ & $1.24 \pm 0.06$ & $1.13 \pm 0.26$ & 3542 & 1094 \\
17 & TOI-1634 b & $4.95 \pm 0.70$ & $1.79 \pm 0.08$ & $0.64 \pm 0.12$ & 3550 & 924 \\
18 & GJ 806 b & $1.92 \pm 0.17$ & $1.33 \pm 0.02$ & $0.73 \pm 0.07$ & 3600 & 937 \\
\enddata
% \tablecomments{Mass, radius, and relative density are listed with symmetric uncertainties and are therefore reported using $\pm$ notation. }
\end{deluxetable*}

\begin{deluxetable*}{lcccccc}
\tablewidth{0pt}
\tablecaption{DACE FGK Planet Sample ($T_{\rm eff}>3800$ K)\label{tab:dace_fgk_sample_full}}
\tablehead{
\colhead{Index} & \colhead{Planet} & \colhead{$M~(M_\oplus)$} & \colhead{$R~(R_\oplus)$} & \colhead{$\rho/\rho_{\rm rock}$} & \colhead{$T_{\rm eff}$ (K)} & \colhead{$T_{\rm eq}$ (K)}
}
\startlongtable
\startdata
1 & HD 260655 b & $2.16 \pm 0.34$ & $1.24 \pm 0.02$ & $1.00 \pm 0.17$ & 3803 & 710 \\
2 & GJ 9827 b & $4.32 \pm 0.34$ & $1.43 \pm 0.08$ & $1.15 \pm 0.21$ & 4236 & 1132 \\
3 & GJ 9827 c & $1.87 \pm 0.38$ & $1.13 \pm 0.06$ & $1.18 \pm 0.30$ & 4236 & 783 \\
4 & TOI-178 c & $4.81 \pm 0.62$ & $1.67 \pm 0.11$ & $0.79 \pm 0.18$ & 4316 & 872 \\
5 & K2-216 b & $8.06 \pm 1.61$ & $1.75 \pm 0.13$ & $1.02 \pm 0.31$ & 4503 & 1101 \\
6 & Kepler-80 d & $3.73 \pm 0.71$ & $1.30 \pm 0.10$ & $1.36 \pm 0.41$ & 4540 & 965 \\
7 & HD 219134 b & $4.78 \pm 0.19$ & $1.60 \pm 0.06$ & $0.89 \pm 0.10$ & 4699 & 1016 \\
8 & HD 219134 c & $4.40 \pm 0.22$ & $1.51 \pm 0.04$ & $1.00 \pm 0.10$ & 4699 & 783 \\
9 & HD 15337 b & $7.26 \pm 0.82$ & $1.70 \pm 0.06$ & $1.02 \pm 0.15$ & 5131 & 990 \\
10 & Kepler-20 b & $9.78 \pm 1.31$ & $1.77 \pm 0.04$ & $1.14 \pm 0.18$ & 5495 & 1190 \\
11 & Kepler-93 b & $4.70 \pm 0.54$ & $1.48 \pm 0.02$ & $1.12 \pm 0.14$ & 5669 & 1134 \\
12 & Kepler-36 b & $3.93 \pm 0.20$ & $1.50 \pm 0.10$ & $0.93 \pm 0.19$ & 5911 & 1082 \\
\enddata
% \tablecomments{Mass, radius, and relative density are listed with symmetric uncertainties and are therefore reported using $\pm$ notation. Equilibrium temperatures are given as central values because no temperature uncertainties were provided in the input tables.}
\end{deluxetable*}

\subsection{Other Supplementary Figures}
\label{sect:sup_figures}

{{This section provides supplementary figures for fitting a mixture model to the CDF of the relative density of planets in different samples. }}

\begin{figure}
\centering
\includegraphics[width=\linewidth]{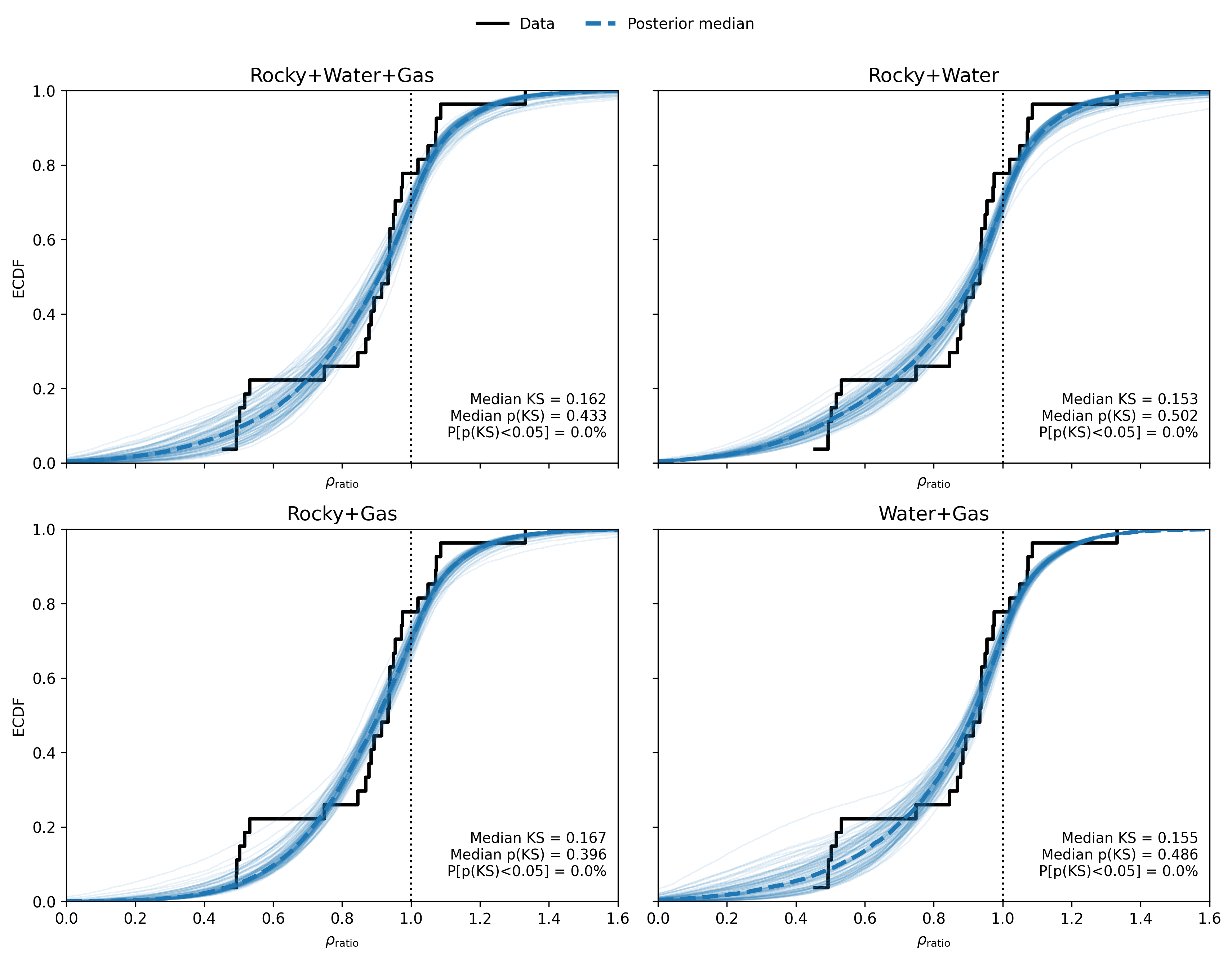}
\caption{Comparison between the observed density-ratio distribution for the {\bf{Luque \& Pall\'e M-dwarf sample}} and posterior realizations using the {\bf{power-law distribution}} under different volatile composition combinations. The black curve shows the cumulative distribution function (CDF) of the data, while the blue curves represent posterior realizations. The dashed blue line indicates the posterior median CDF. }
\label{fig:luque_cdf_powerlaw}
\end{figure}

\begin{figure}
\centering
\includegraphics[width=\linewidth]{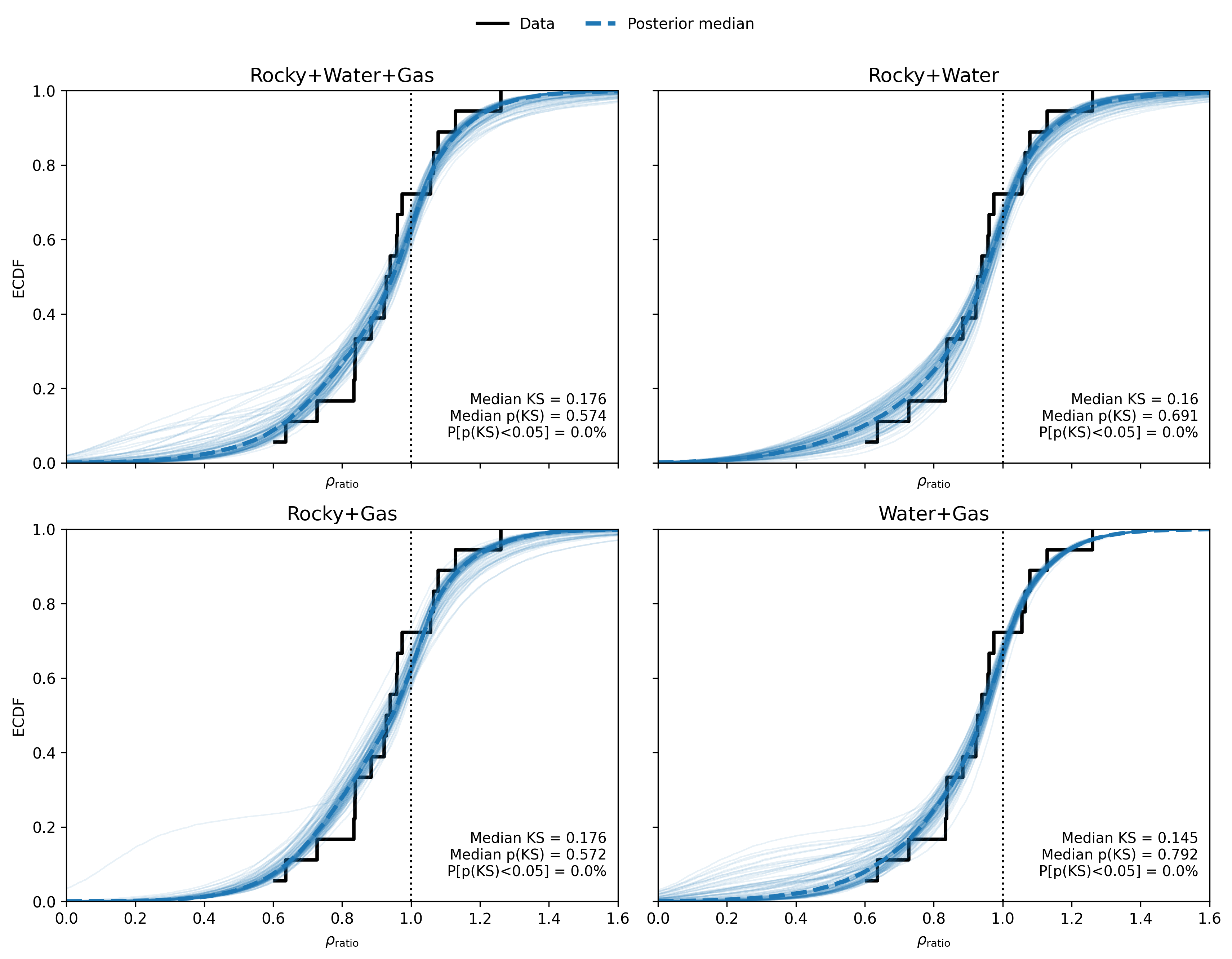}
\caption{Comparison between the observed density-ratio distribution for the {\bf{DACE M-dwarf sample}} and posterior realizations using the {\bf{power-law distribution}} under different volatile composition combinations. The black curve shows the cumulative distribution function (CDF) of the data, while the blue curves represent posterior realizations. The dashed blue line indicates the posterior median CDF.}
\label{fig:dace_m_cdf_power_law}
\end{figure}

\begin{figure}
\centering
\includegraphics[width=\linewidth]{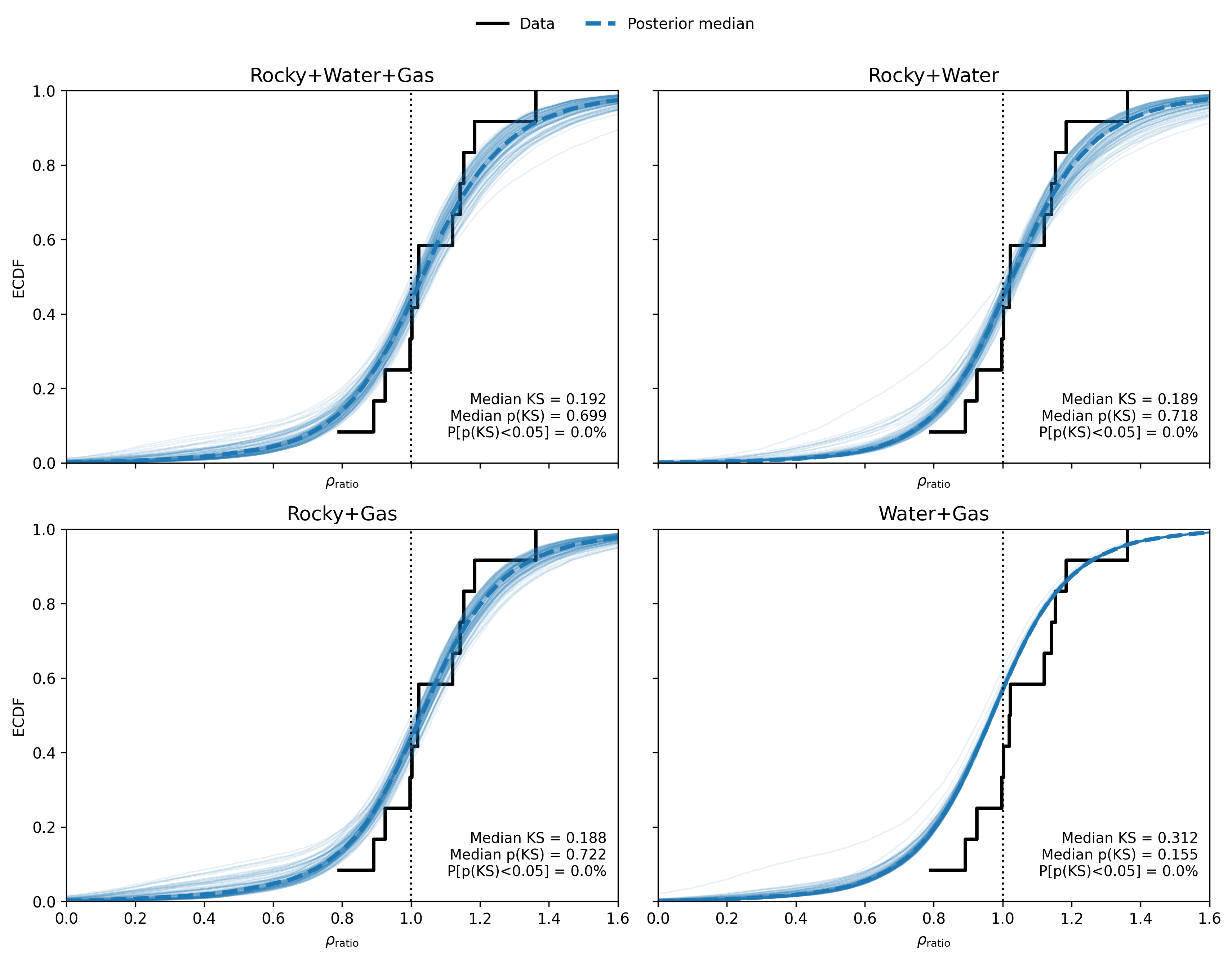}
\caption{Same as Figure~\ref{fig:dace_m_cdf_power_law}, but for the {\bf{DACE FGK sample}} and posterior realizations using the {\bf{power-law distribution}}.}
\label{fig:dace_fgk_cdf_power_law}
\end{figure}

% \begin{figure}
% \centering
% \includegraphics[width=\linewidth]{posterior_cdf_2x2_globalx.png}
% \caption{Comparison between the observed density-ratio distribution for the {\bf{Luque \& Pall\'e M-dwarf sample}} and posterior predictive models using the {\bf{log-normal distribution}} under different volatile composition combinations. The black curve shows the cumulative distribution function (CDF) of the data, while the blue curves represent posterior predictive realizations. The dashed blue line indicates the posterior median CDF. }
% \label{fig:luque_cdf}
% \end{figure}

% \begin{figure}
% \centering
% \includegraphics[width=\linewidth]{posterior_cdf_2x2_globalx.dace_m.png}
% \caption{Comparison between the observed density-ratio distribution for the {\bf{DACE M-dwarf sample}} and posterior predictive models using the {\bf{log-normal distribution}} under different volatile composition combinations. The black curve shows the cumulative distribution function (CDF) of the data, while the blue curves represent posterior predictive realizations. The dashed blue line indicates the posterior median CDF.}
% \label{fig:dace_m_cdf}
% \end{figure}

% \begin{figure}
% \centering
% \includegraphics[width=\linewidth]{posterior_cdf_2x2_globalx.dace_fgk.png}
% \caption{Same as Figure~\ref{fig:dace_m_cdf}, but for the {\bf{DACE FGK sample}} and a posterior predictive models using the {\bf{log-normal distribution}}. }
% \label{fig:dace_fgk_cdf}
% \end{figure}

\end{CJK*}

\end{document}